\documentclass[pdftex,twocolumn,epjc3]{svjour3}          

\RequirePackage[T1]{fontenc}

\smartqed  
\RequirePackage{multirow}
\RequirePackage{graphicx}
\RequirePackage{mathptmx}      
\RequirePackage{flushend}
\RequirePackage[numbers,sort&compress]{natbib}
\RequirePackage[colorlinks,citecolor=blue,urlcolor=blue,linkcolor=blue]{hyperref}

\usepackage{graphicx}  
\usepackage{dcolumn}   
\usepackage{bm}        
\usepackage{amssymb}   
\usepackage{feynmf}
\usepackage{hyperref}
\usepackage{slashed}
\usepackage{multirow}
\usepackage{color}
\usepackage{array}
\usepackage{caption}
\usepackage{subcaption}
\usepackage{amsmath}
\usepackage{booktabs}
\usepackage{multirow}
\usepackage{booktabs}
\usepackage{soul}
\usepackage{xspace}
\usepackage{import}
\usepackage{float}
\usepackage{microtype}
\usepackage{dblfloatfix}

\newcommand{\asterisknewline}[1]{%
}

\newcommand{\hg}{HGPflow\xspace}
\newcommand{\pf}{PPflow\xspace}
\newcommand{\pd}{Pandora\xspace}
\newcommand{\pt}{$p_\mathrm{T}$\xspace}
\newcommand{\met}{$p_\mathrm{T}^\mathrm{miss}$\xspace}
\newcommand{\ttbar}{$t\overline{t}$\xspace}

\begin{document}

\title{\hg: Extending Hypergraph Particle Flow to Collider Event Reconstruction}

\author{
    Nilotpal Kakati \thanksref{wis_add, e1}
    \and Etienne Dreyer \thanksref{wis_add}
    \and Anna Ivina \thanksref{wis_add}
    \and Francesco Armando Di Bello \thanksref{gen_add}
    \and Lukas Heinrich \thanksref{tum_add}
    \and Marumi Kado \thanksref{mpi_add}
    \and Eilam Gross \thanksref{wis_add}
}

\thankstext{e1}{e-mail: nilotpal.kakati@weizmann.ac.il}

\institute{
    Weizmann Institute of Science \label{wis_add}
    \and INFN and University of Genova \label{gen_add}
    \and Technical University of Munich \label{tum_add}
    \and Max Planck Institute for Physics \label{mpi_add}
}

\date{Received: date / Accepted: date}

\maketitle

\begin{abstract}

In high energy physics, the ability to reconstruct particles based on their detector signatures is essential for downstream data analyses. A particle reconstruction algorithm based on learning hypergraphs (HGPflow) has previously been explored in the context of single jets. In this paper, we expand the scope to full proton-proton and electron-positron collision events and study reconstruction quality using metrics at the particle, jet, and event levels. Instead of passing entire events through HGPflow, we train it on smaller partitions for scalability and to avoid potential bias from long-range correlations related to the physics process. We demonstrate that this approach is feasible and that on most metrics, HGPflow outperforms both traditional particle flow algorithms and a machine learning-based benchmark model. 

\end{abstract}

\section{Introduction} \label{sec:introduction}
\begin{figure*}[!ht]
    \centering
    \includegraphics[width=\textwidth]{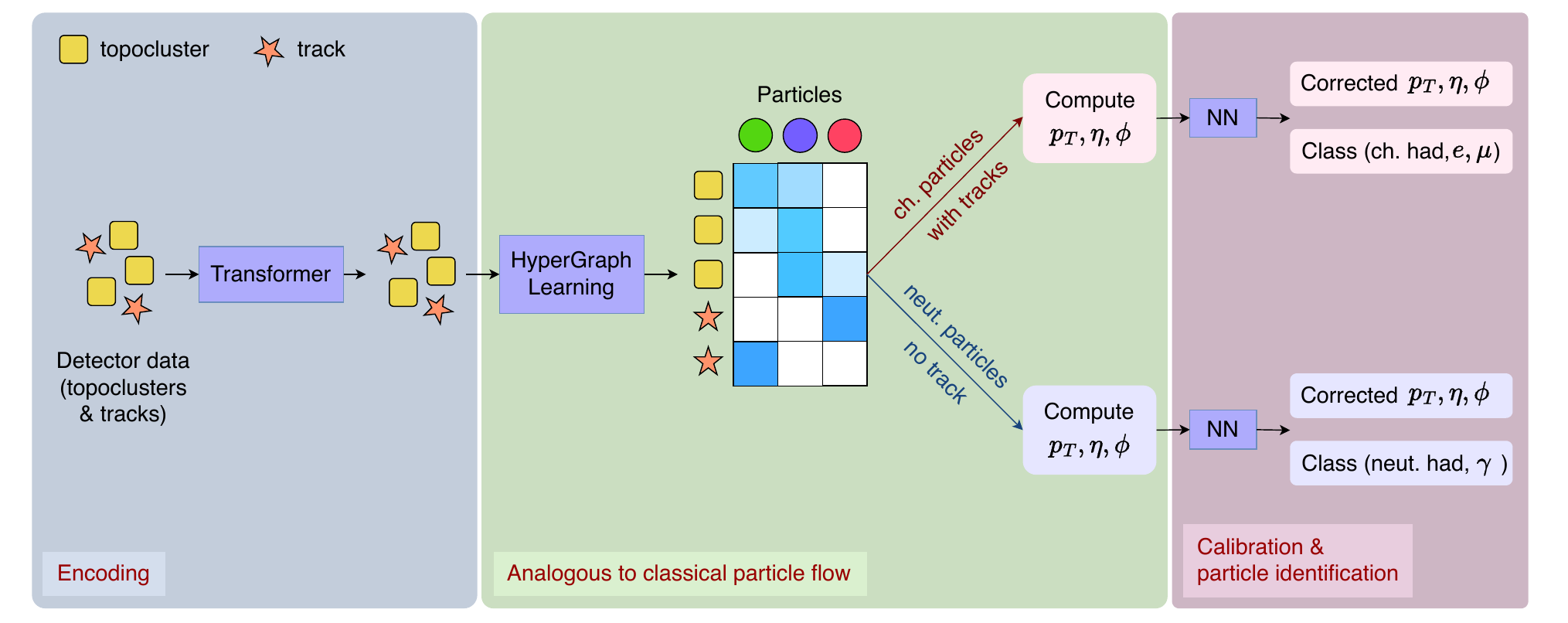}
    \caption{Schematic showing the three main blocks of the HGPflow algorithm.}
    \label{fig:hgpflow_diagram}
\end{figure*}

Collider experiments have enabled historic breakthroughs in the field of particle physics and continue to drive progress on fundamental questions. To study the interactions underlying particle collisions, detectors measure outgoing particles via the energy they deposit in a series of ``hits'' as they traverse concentric arrays of sensors. For general-purpose detectors at the Large Hadron Collider (LHC) \cite{lhcMachine}, tracker and calorimeter hits comprise the bulk of information for pattern recognition algorithms that reconstruct the particles produced in each collision. Low-level reconstruction algorithms are designed to cluster hits in the tracking volume \cite{atlastracking,cmstracking} and calorimeter \cite{atlasclustering,cmsclustering} separately to form tracks and calorimeter clusters. High-level \textit{particle flow} algorithms \cite{ALEPH:1994ayc,ATLAS:2017ghe,sirunyan2017particle,Thomson:2009rp} then attempt to combine the track and cluster information in an optimal way to form reconstructed objects. Ideally, these particle flow objects mimic the true particles, offering a universal starting point for higher-level reconstruction, including jets, physics observables, and event topology.

In a given detector, the difficulty of reconstructing particles depends on intrinsic factors such as their momentum and type, as well as their local environment. While charged particles are generally well-modeled by track reconstruction, at higher momenta, calorimeter clusters associated with the track can significantly improve the measurement. In contrast, reconstructing long-lived neutral hadrons and photons relies entirely on calorimetry, which is more straightforward for electromagnetic showers than for hadronic showers. Hadronic showers exhibit large spatial fluctuations and dissipate significant fractions of their energy in forms that are invisible to most calorimeters. Moreover, a simple one-to-one relationship between calorimeter clusters and particles is often degraded by limited calorimeter granularity and the effects of noise suppression.

These challenges are further exacerbated in dense environments like hadronic jets, where calorimeter signatures from multiple particles frequently overlap. A substantial component of jet energy resolution thus stems from \textit{confusion}~-- the difficulty in correctly associating calorimeter clusters with tracks and distinguishing hadronic and electromagnetic showers. Traditional particle flow algorithms address this by using parameterized energy profiles for each track to associate nearby calorimeter cells or to identify their neutral particle origin. However, the high-dimensional nature of the input data and sensitivity to environmental factors motivate a multivariate approach as a solution.

In recent years, several efforts have explored modernizing the particle flow concept with the application of deep learning (DL) techniques \cite{DiBello:2020bas,Kieseler_2020,pata2021mlpf,pflow,qasim2022end,Pata:2023rhh}. From a DL perspective, particle reconstruction is a prediction task using the features associated with the set of detector hits\footnote{Depending on the approach, ``hits'' can refer to tracks, calorimeter cells, and calorimeter clusters.} as an input to predict a ground-truth set of particles, including their class, direction, and momentum. The permutation invariance of the input and output sets motivates the use of graph neural networks \cite{battaglia2018relationalinductivebiasesdeep} and transformers \cite{vaswani2023attentionneed}, as in various other applications to high energy physics \cite{Shlomi_2020, DeZoort:2023vrm,Qasim:2019otl,Qu:2022mxj,Finke:2023veq}. The existing algorithms for particle reconstruction using DL can be compared based on how the set-to-set task is defined in each case.

In object condensation~\cite{Kieseler_2020,qasim2022end}, the network learns to cluster detector hits based on parent particle identity in a representation space guided by attractive and repulsive potentials as losses. Additional loss terms supervise the particle properties associated with each hit and encourage the promotion of some hits to serve as centroids in each cluster. The Machine Learning Particle Flow (MLPF) approach~\cite{pata2021mlpf,Pata:2023rhh} is similar in that particle properties are supervised for each hit, but differs in that the cardinality of particles is obtained by classifying hits as either belonging to a valid particle type or else belonging to the ``reject'' category. The first algorithm proposed in \cite{pflow} -- Transformer Set Prediction Network with Slot Attention (TSPN-SA) -- links hits and particles through an attention matrix in an unsupervised manner. Finally, Hypergraph Particle Flow (HGPflow)~\cite{pflow} builds on the idea of a learnable matrix relating the input and output sets but defines this matrix in physical terms and promotes it to a supervised training objective.

In this paper, we extend the work of \cite{pflow} by applying HGPflow to full collision events and evaluating a variety of performance metrics for different physics processes. Section~\ref{sec:algorithms} describes the HGPflow algorithm along with other algorithms used as benchmarks, including MLPF. In Sec.~\ref{sec:event_partitioning}, we introduce event partitioning as a strategy to train DL algorithms on full collision events which is computationally efficient and avoids potential bias from long-range physical correlations. We test HGPflow in both an LHC-like scenario, simulated using COCOA \cite{COCOA}, and a future $e^+e^-$ collider scenario, simulated using the Compact Linear Collider (CLIC) detector model \cite{CLICdp:2017vju, CLICdp:2018vnx}. The training and testing datasets are described in Sec.~\ref{sec:dataset}, the training target and procedure in Sec.~\ref{sec:target}, and the performance results in Sec.~\ref{sec:results}. 

\section{Algorithms} \label{sec:algorithms}


\subsection{\hg algorithm}

\begin{figure}[!h]
    \centering
    \includegraphics[width=0.4\textwidth]{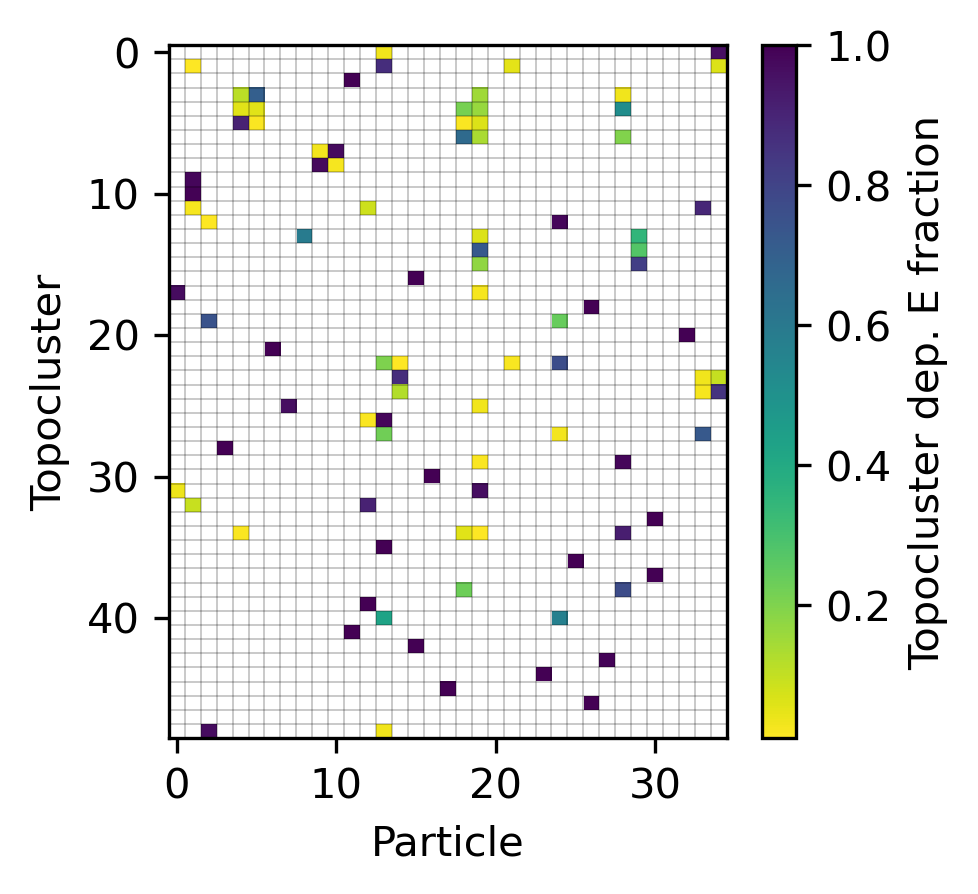}
    \caption{An example of the energy-based incidence matrix relating calorimeter clusters (``topoclusters'') to particles for a dijet event from the test dataset. Entries for tracks are not shown.}
    \label{fig:ex_incidence}
\end{figure}

The problem of correctly assigning energy deposits in the calorimeter to particles can be naturally formulated as a hypergraph learning task, as first proposed in \cite{pflow}. A hypergraph is defined by a set of nodes (vertices) $\mathcal{V}=\{V_i; i=1\dots N\}$, a set of hyperedges $\mathcal{E}=\{E_a; a=1\dots K\}$, and an incidence matrix $I^{N\times K}$ describing the relational structure between the two sets. The entry $I_{ia}$ describes the weight with which node $i$ is a member of hyperedge $a$. As shown in Fig.~\ref{fig:hgpflow_diagram}, the central objective of the HGPflow algorithm is to predict the incidence matrix relating the set of tracks and calorimeter clusters\footnote{These relations could also be defined at the level of cells, as in \cite{Kakati:2024bjf}.} ($\mathcal{V}$) to the set of particles, represented by hyperedges ($\mathcal{E}$). The relationship between tracks and particles is represented by binary entries in the incidence matrix. For calorimeter clusters, the incidence matrix entry $I_{ia}$ denotes the fraction of a cluster $i$'s true energy that it received from particle $a$:
\begin{equation} \label{eq:incidence}
    I_{ia} = \frac{E_{ia}}{\sum\limits_{\mathrm{particles}\ b}E_{ib}} = \frac{E_{ia}}{E_i}
\end{equation}
An example incidence matrix is shown in Fig.~\ref{fig:ex_incidence}. The advantage of formulating particle reconstruction like this is that energy deposits in the calorimeter are assigned to particles in a manner that conserves energy by construction. If a fraction $f$ of a given calorimeter cluster is associated with one particle, only $1-f$ of its energy remains to be associated with other particles. In HGPflow, this is achieved by applying a SoftMax operation~($z_a \rightarrow e^{z_a} / \Sigma_{b} e^{z_b}$) to normalize the predicted incidence matrix entries for each cluster. The physical constraint thus serves as an inductive bias, enhancing the performance and interpretability of HGPflow predictions.

The first stage of \hg produces a hidden representation of tracks and clusters by first encoding their features in the same high-dimensional space and then updating them in the transformer encoder shown in Fig.~\ref{fig:hgpflow_diagram}. Since the true number of particles is unknown \textit{a priori}, the number of candidate hyperedges, $K$, is set to a value large enough to serve as an upper bound on the number of particles per example.
The incidence matrix is initialized such that tracks are uniquely assigned to hyperedges via fixed one-hot encodings. Then, for each cluster, the network predicts a vector of length $K$ obtained through an iterative refinement sequence comprising 12 update blocks, following \cite{zhang2021recurrently}. As in \cite{pflow}, we employ backpropagation through time, meaning that gradients are only computed for two sequences of 4 consecutive update blocks each that are randomly selected from the 12 total iterations. The Kullback-Leibler divergence is used as the loss function, with the Hungarian algorithm~\cite{kuhn1955hungarian} employed prior to backpropagation to determine the assignment between predicted and target hyperedges that minimizes the loss. Finally, to determine particle cardinality, HGPflow predicts an \textit{indicator score} for each hyperedge determining whether it corresponds to a real particle or not. During inference, hyperedges are dropped unless they contain a track or have an indicator score above a certain threshold.


If the incidence matrix is learned properly, the energy and angular coordinates of each neutral particle can already be approximated based on its association with the energy deposits $E_i$ in the calorimeter, as follows: 
%
\begin{align}
    \hat{E}_a &= \sum_i I_{ai} \cdot E_i \label{eq:proxy1} \\
    \{ \hat{\eta}_a, \hat{\phi}_a\} &= \frac{1}{\hat{E}_a}\sum_i I_{ai} \cdot E_i \cdot \{ \eta_i, \phi_i\} =\sum_i \tilde{I}_{ai} \cdot \{ \eta_i, \phi_i\}, \label{eq:proxy2}
\end{align}
where $\tilde{I}_{ai} = E_{ia} / E_a$ denotes a \textit{dual} incidence matrix which is normalized column-wise rather than row-wise. We apply Eqs.~\ref{eq:proxy1}-~\ref{eq:proxy2} to the predicted incidence matrix to construct a set of ``proxy'' particles. For hyperedges containing a track, we overwrite the proxy properties with those of the track. The creation of proxy particles is shown on the right-hand side of the second stage in Fig.~\ref{fig:hgpflow_diagram}.

The accuracy of proxy particles is mainly limited by track resolution and calorimeter characteristics such as granularity, upstream material interactions, electronic noise levels, and sampling fraction. The final objective of HGPflow, shown in the third stage of Fig.~\ref{fig:hgpflow_diagram}, is thus to recover the true particle properties as a learned correction to the proxy properties. Each particle is corrected using as input its proxy kinematics, node features weighted by the predicted incidence matrix, and the hidden representation of the corresponding hyperedge. For the datasets considered in this paper, correcting angular coordinates had negligible impact, so only transverse momentum and energy are corrected for charged particles and neutral particles, respectively. Furthermore, the CLIC calorimeter information proved good enough that no corrections were applied to the proxy particles obtained directly from the incidence matrix prediction.
Classification of reconstructed particles is also performed at this stage. Hyperedges with an associated track are classified as either charged hadron, electron, or muon, while the rest are split into photon and neutral hadron categories.

Compared to the version in \cite{pflow}, the updated HGPflow model benefits from a more modern DL architecture. The node encoding network utilizes the diffusion transformer (DiT) architecture proposed in \cite{Peebles2022ScalableDM}. Compared to the original hypergraph update sequence proposed in \cite{zhang2021recurrently}, we introduced a transformer in the hyperedge update to encode context with self-attention. Transformer operations are performed with \textsc{FlashAttention-2} \cite{dao2022flashattentionfastmemoryefficientexact,dao2023flashattention2fasterattentionbetter} for improved memory management and computational speedup. Refer to Sec.~\ref{sec:target} for the target definition, training procedure, and hyperparameters for HGPflow.

\subsection{MLPF algorithm}\label{sec:mlpf}

As a DL model for particle reconstruction trained on full events, the MLPF algorithm \cite{pata2021mlpf,Pata:2023rhh} provides the closest comparison to our approach. Calorimeter clusters and tracks make up the set of inputs that MLPF uses to predict the set of target particles. Each target particle is associated with a unique input. A graph convolutional network \cite{kipf2017semisupervised} is used to update the hidden representation of each input based on its local context. This representation is used to predict the class, charge, and four-momentum of the target particles. The network is trained to classify each input to match the class of its assigned particle, while inputs without assignment are classified as unassigned and rejected. The latest version of MLPF uses Locality Sensitive Hashing (LSH) \cite{kitaev2020reformerefficienttransformer} to divide the inputs into bins based on their proximity in a learned feature space. Graph connectivity is restricted to the LSH bins, allowing information exchange that remains computationally scalable.
The authors of \cite{Pata:2023rhh} experimented with a model that operates on the level of calorimeter cells, but ultimately found that using calorimeter clusters is more feasible and leads to better performance. More details about the target definition and model configuration we used for MLPF can be found in Sec.~\ref{sec:target}.

\subsection{\pf algorithm}

To represent traditional reconstruction approaches, we use a parameterized particle flow (\pf) algorithm as a baseline. The PPflow algorithm is designed to improve jet momentum resolution by combining tracks and calorimeter clusters while avoiding double counting. This is accomplished by deriving templates of the energy distribution from single charged pions in the calorimeter. In jets, where hadronic showers from multiple charged and neutral sources overlap, these pion energy templates can then be subtracted from calorimeter deposits to infer the contributions from neutral particles. Summing the energy associated with the track with the energy of the residual neutral components provides a more optimal use of the calorimeter and tracker information.

\subsection{\pd algorithm}

\pd \cite{Thomson:2009rp} is a highly sophisticated particle flow algorithm developed for future linear colliders and recently applied to the CLIC detector~\cite{CLICdp:2017vju}. Exploiting the high granularity of the CLIC calorimeter, \pd clusters cells within cones aligned with the cluster direction, working outward from the innermost calorimeter layer, where track projections are used to seed clusters when present. Clusters without an associated track are identified as photons if they satisfy specific criteria; otherwise, they are considered for merging with neighboring track-associated clusters. Subsequently, clusters associated with a track are iteratively re-clustered to find a configuration where the track momentum is compatible with the associated calorimeter energy. Particle flow objects from \pd are classified as either photon, electron, muon, charged hadron, or neutral hadron using a simple particle identification scheme. Software compensation is employed in \pd to enhance jet energy resolution by identifying electromagnetic sub-showers embedded inside of hadronic showers and calibrating each separately \cite{PandoraSoftwareCompensation}. These careful design choices complement the excellent CLIC detector performance to achieve a strong correspondence between particle flow objects from \pd and truth particles. Performance studies of \pd in the CLIC context demonstrated an impressive 3.5\% jet energy resolution across a wide energy spectrum \cite{Linssen:1425915}.

\section{Event partitioning}\label{sec:event_partitioning}
Scaling up the task of particle reconstruction from single jet examples used in \cite{pflow} to full collision events introduces two potential challenges. The first challenge is computational, given the increased dimensionality of the input space. A second challenge arises because training on full events opens the possibility of learning global features generated by the hard-scatter process. In Sec.~\ref{sec:scaling} and Sec.~\ref{sec:correlations}, we argue that these challenges can be addressed by splitting events into smaller partitions and propose an event partitioning scheme in Sec.~\ref{sec:MS}.

\subsection{Memory scaling}\label{sec:scaling}

Memory requirements for transformer architectures are sensitive to the maximal dimension $N$ of the input nodes, which is in the hundreds for our training samples. Since the incidence matrix has the shape $(N \times K)$, it scales less than quadratically in $N$, because the upper bound $K$ on the number of target particles is typically less than $N$. The main memory consumption in HGPflow arises from the transformer encoder (the first stage in Fig.~\ref{fig:hgpflow_diagram}), which scales quadratically with $N$. For relatively clean collision environments, this remains feasible to allocate for a full event when working at the level of tracks and calorimeter clusters. However, for high-multiplicity $pp$ collisions including pileup, a more efficient approach is required. Operating at the level of cells instead of clusters can similarly increase $N$ by one to two orders of magnitude, depending on calorimeter granularity.

Although the scalability of transformers has been thoroughly studied with promising alternative solutions such as~\cite{tay2022efficienttransformerssurvey,Sun2023-qd,longnet2023}, our approach takes a different direction by scaling down the problem instead of scaling up the solution. Partitioning the event and restricting training and inference to the smaller partitions makes the memory overhead tuneably small.

In~\cite{Pata:2023rhh}, local sensitivity hashing (LSH) was employed for manageable memory scaling. Although LSH is highly efficient for retrieving similar items in large databases, it offers no guarantee that the resulting clusters are strictly local, which is a requirement for a meaningful incidence matrix. Therefore, we opt for a clustering approach that meets this requirement by operating directly in the angular space of the detector.

\subsection{Long-range correlations}\label{sec:correlations}

\begin{figure}[!ht]
    \centering
    \includegraphics[width=0.48\textwidth]{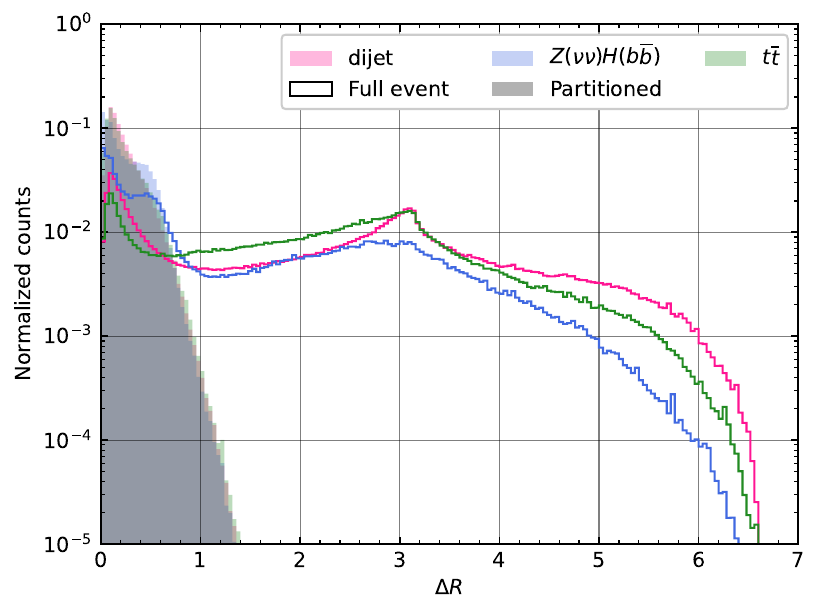}
    \caption{Histograms of the angular separation between pairs of calorimeter clusters before (empty) and after (filled) partitioning for three different physics processes simulated in COCOA. Each pair enters the histogram weighted by the product of the two energies.}
    \label{fig:correlation}
\end{figure}

Correlations among the energy and spatial coordinates of outgoing particles carry insights about the hard scatter process, including its topology, momentum balance, and interaction type. 
Figure \ref{fig:correlation} shows correlations in the angular distribution of calorimeter clusters, weighted by energy, revealing a dependence on physics process. The peak at $\Delta R = \pi$ reveals back-to-back correlation, and is notably reduced for the boosted $Z(\nu\nu)H(b\overline{b})$ events, as expected. These events, on the other hand, feature a bump in the region $0 < \Delta R < 1$ consistent with the opening angle between the decay products.

For particle reconstruction to function universally, it is best approached as a local problem, agnostic to long-range correlations. If reconstructed outputs provide information that is biased toward the physics process used to derive them, they no longer serve their purpose as generic inputs for high-level analyses.  Since DL models naturally exploit all available features, they are free to use both local and global correlations present in the training data. Thus, when using DL for particle reconstruction, special care is needed to avoid inadvertently learning non-local, process-specific features.

Event partitioning is a straightforward way to remove long-range correlations. In this approach, the partition determines the local scope of the question and, hence, of the answer. Figure~\ref{fig:correlation} demonstrates that correlations beyond $\Delta R \simeq 1$ are removed after event partitioning. In Sec.~\ref{sec:locality} we show that long-range correlations can get encoded in a DL model during training and influence its physics performance.

\subsection{Mean Shift clustering}\label{sec:MS}

\begin{figure}[!th]
    \centering
    \includegraphics[width=0.35\textwidth]{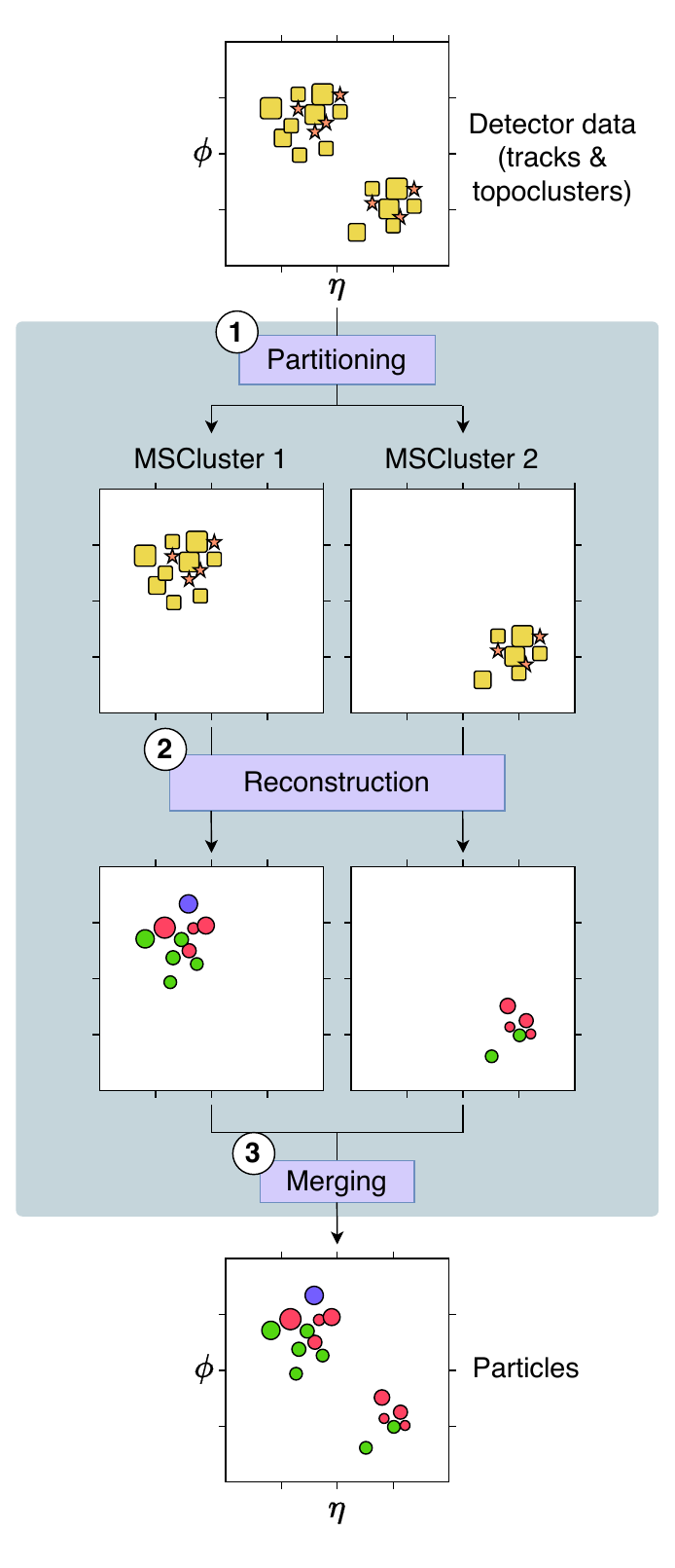}
    \caption{The event partitioning procedure using Mean Shift clustering in the $\eta$-$\phi$ plane.}
    \label{fig:hgpflow_demo}
\end{figure}

Our event partitioning approach is illustrated in Fig.~\ref{fig:hgpflow_demo}. In the first step, an algorithm divides the full set of tracks and calorimeter clusters into smaller partitions. Reconstruction is then performed for each partition in parallel. Finally, the predictions from each partition are merged, resulting in a set of reconstructed particles for the full event.

The choice of algorithm to perform the event partitioning in this paper is guided by a few desired properties. First, a unique assignment of \textit{all} detector tracks and calorimeter clusters to disjoint partitions is necessary to avoid the need for overlap removal on the set of predicted particles. The algorithm also needs to be suitable for a variable number of input nodes and able to cope with significant variations in local density. While jet clustering algorithms based on sequential recombination are a natural choice for collision events, they are not designed to create clusters outside of jets. A handful of popular, generic algorithms also seem less than suitable: $k$-means clustering requires assuming the number ($k$) of clusters beforehand, DBScan~\cite{ester1996density} is intended for clusters of similar high density, and OPTICS~\cite{optics} can leave some nodes clustered.

The Mean Shift (MS)~\cite{meanshift1,meanshift2} clustering algorithm is a straightforward choice that fulfills the above criteria. The MS algorithm works by iteratively shifting input nodes in the direction of higher data density. Given a distance metric, the process starts by defining a neighborhood around each node of size $h$ called the \textit{bandwidth}. After computing the centroid of each neighborhood as the mean over the nodes inside, the reference node is shifted in the direction of the centroid. This process repeats until convergence when centroids coalesce into stationary peaks of high density. Finally, the bandwidth is used to determine whether nodes belong to one cluster or another. This procedure requires minimal assumptions about the number of clusters or their density distribution. Although it works best for spherical clusters, it behaves well for nodes in relatively sparse intermediate regions.

We used a modified version of the \textsc{Scikit-Learn}~\cite{sklearn} implementation of Mean Shift, with the default Cartesian distance metric replaced with $\Delta R$ to take into account the periodicity in $\phi$. For clusters, the barycenter $\eta$ and $\phi$ are used, whereas, for tracks the extrapolated position at the calorimeter face is used. The bandwidth $h$ is set to 0.4 and 0.3 in our results on $pp$ and $e^+e^-$ events, respectively. This configuration gave good results in limited studies but should be properly optimized in future work.

The partitioning approach introduces edge effects when two or more energy deposits from a single particle get split among multiple partitions. In such cases, each partition holds partial information about the same particle. If the target in each MS cluster is the original particle, double counting will arise once predictions are merged. In practice, this predominantly affects hadrons within jets. We deal with this effect in two separate cases, as follows. Case 1: if the split particle has no associated track, we create two separate target particles that have their kinematics recomputed based on the fraction of energy in each MS cluster. If both targets are successfully reconstructed, the original energy will be recovered upon merging, although there will be two neutral particles rather than one. Case 2: if the split particle has a track, we keep the original target particle only in the MS cluster that contains the track. Energy deposits from the original particle that end up in other MS clusters are then associated with a modified target particle called a \textit{residual} particle. Since they have actual energy deposits, residual particles participate in the energy assignment of the target incidence matrix. However, residual particles are assigned a target indicator score of zero, training HGPflow to reject them along with non-existent hyperedges. 

\begin{figure}[H]
    \centering
    \includegraphics[width=0.48\textwidth]{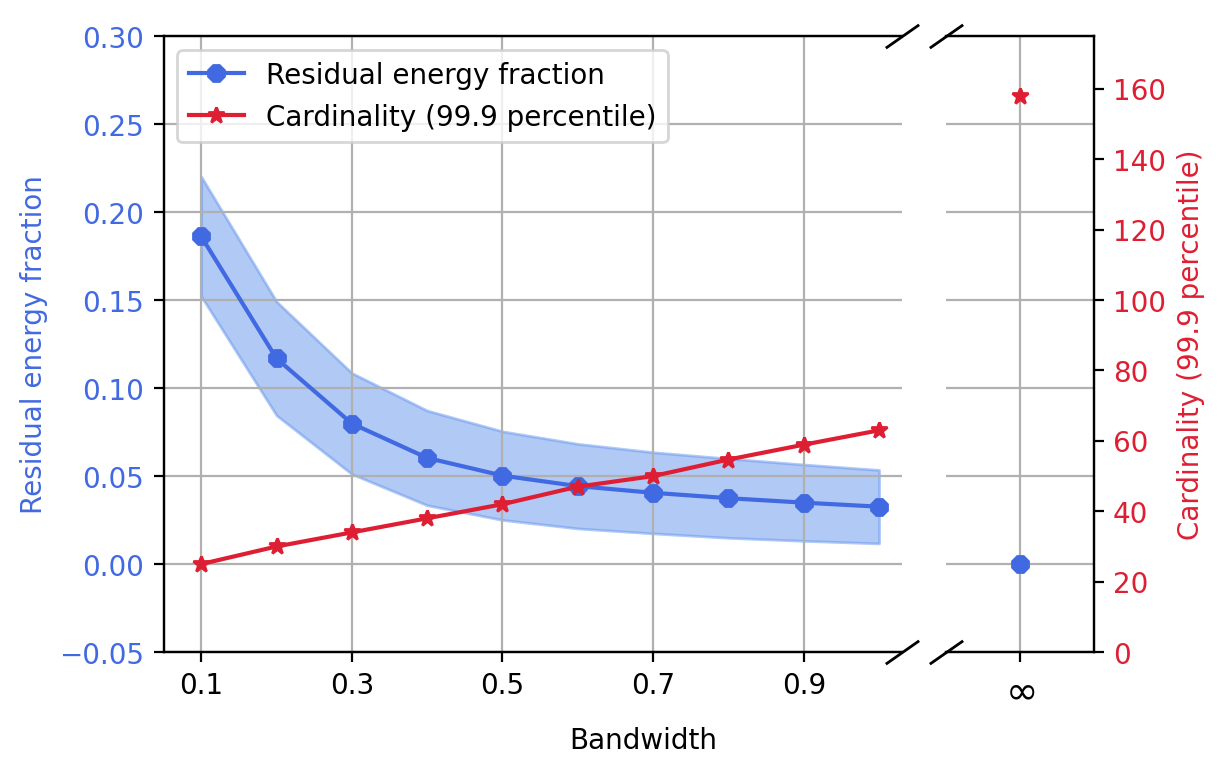}
    \caption{The impact of bandwidth on residual energy fraction per event and particle cardinality per MS cluster. The band on residual energy indicates the interquartile range around the median, denoted by the markers.}
    \label{fig:bandwidth}
\end{figure}

Figure \ref{fig:bandwidth} shows that the fraction of energy attributed to residual particles decreases for larger values of bandwidth. Conversely, the upper bound on the number of particles per MS cluster is directly proportional to the bandwidth. These trends demonstrate the interplay between edge effects and computational cost when deciding the characteristic size of event partitions.



\section{Datasets}
\label{sec:dataset}
\begin{figure*}[!hb]
    \centering
    \begin{subfigure}[c]{0.297\textwidth}
        \includegraphics[width=\textwidth]{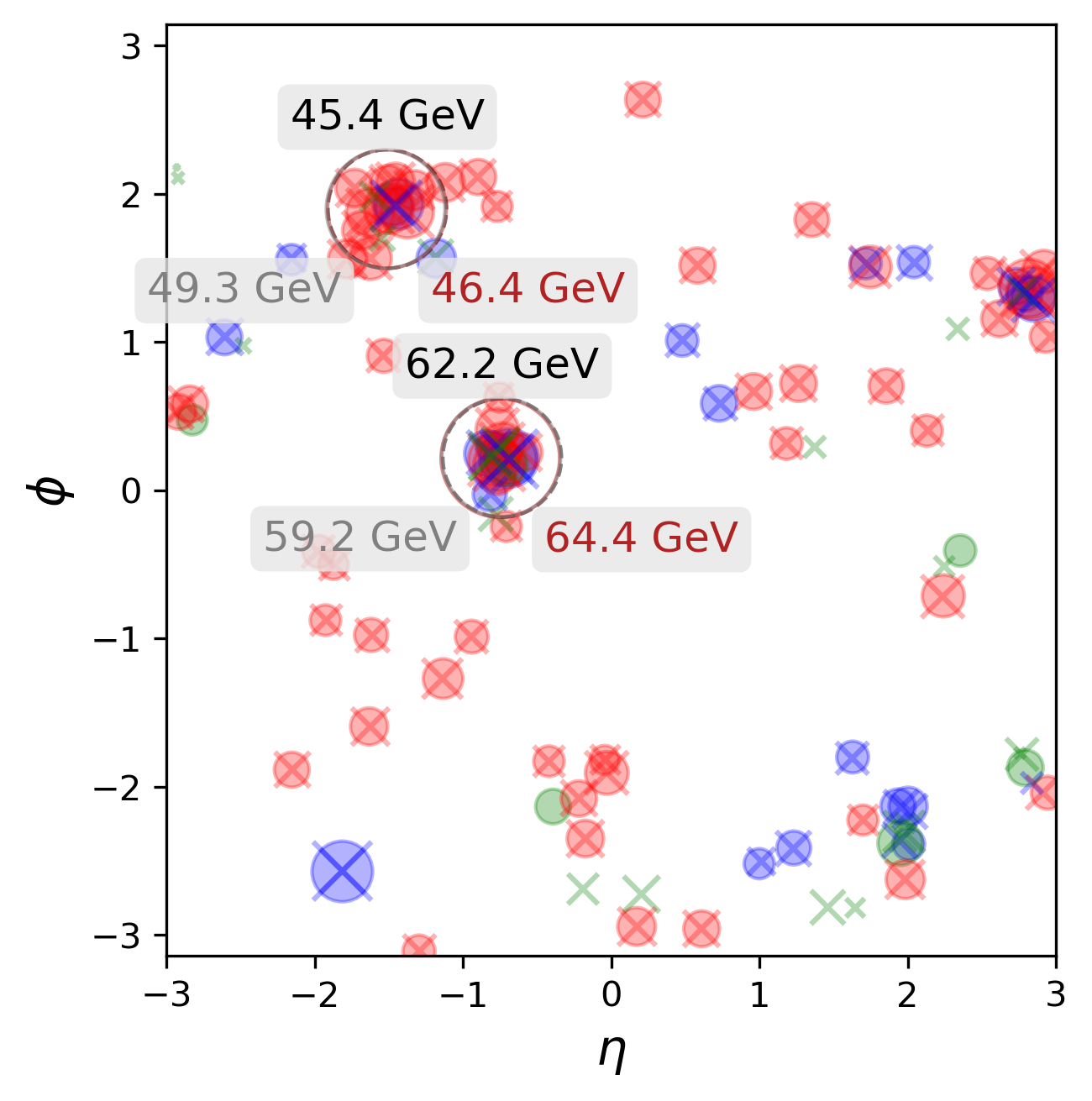}
        \caption{Dijet}
        \label{fig:zhbb_mass_residual}
    \end{subfigure}
    \begin{subfigure}[c]{0.297\textwidth}
        \includegraphics[width=\textwidth]{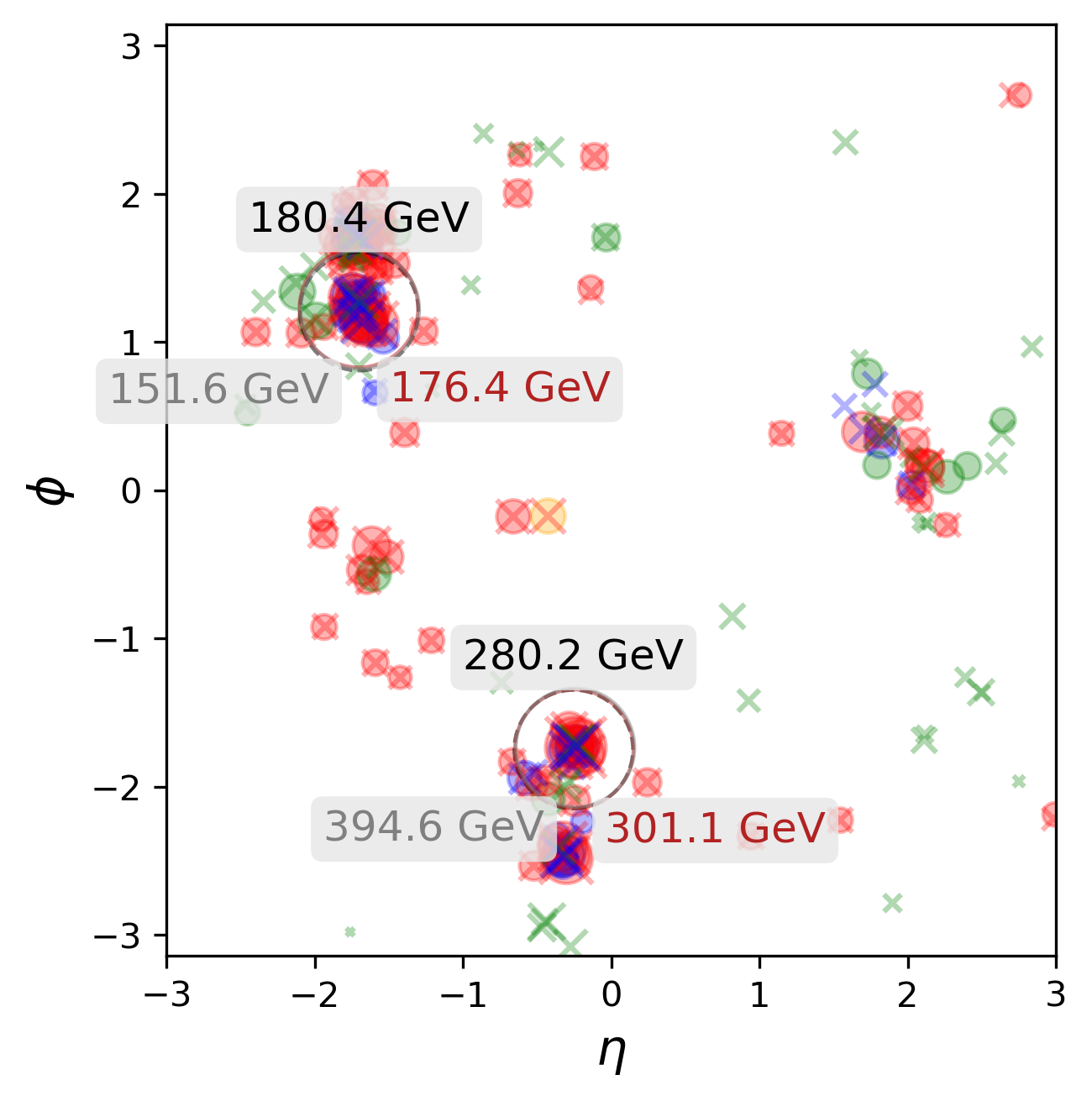}
        \caption{$t\bar{t}$}
        \label{fig:zhbb_c2}
    \end{subfigure}
    \begin{subfigure}[c]{0.397\textwidth}
        \vspace{1mm}
        \includegraphics[width=\textwidth]{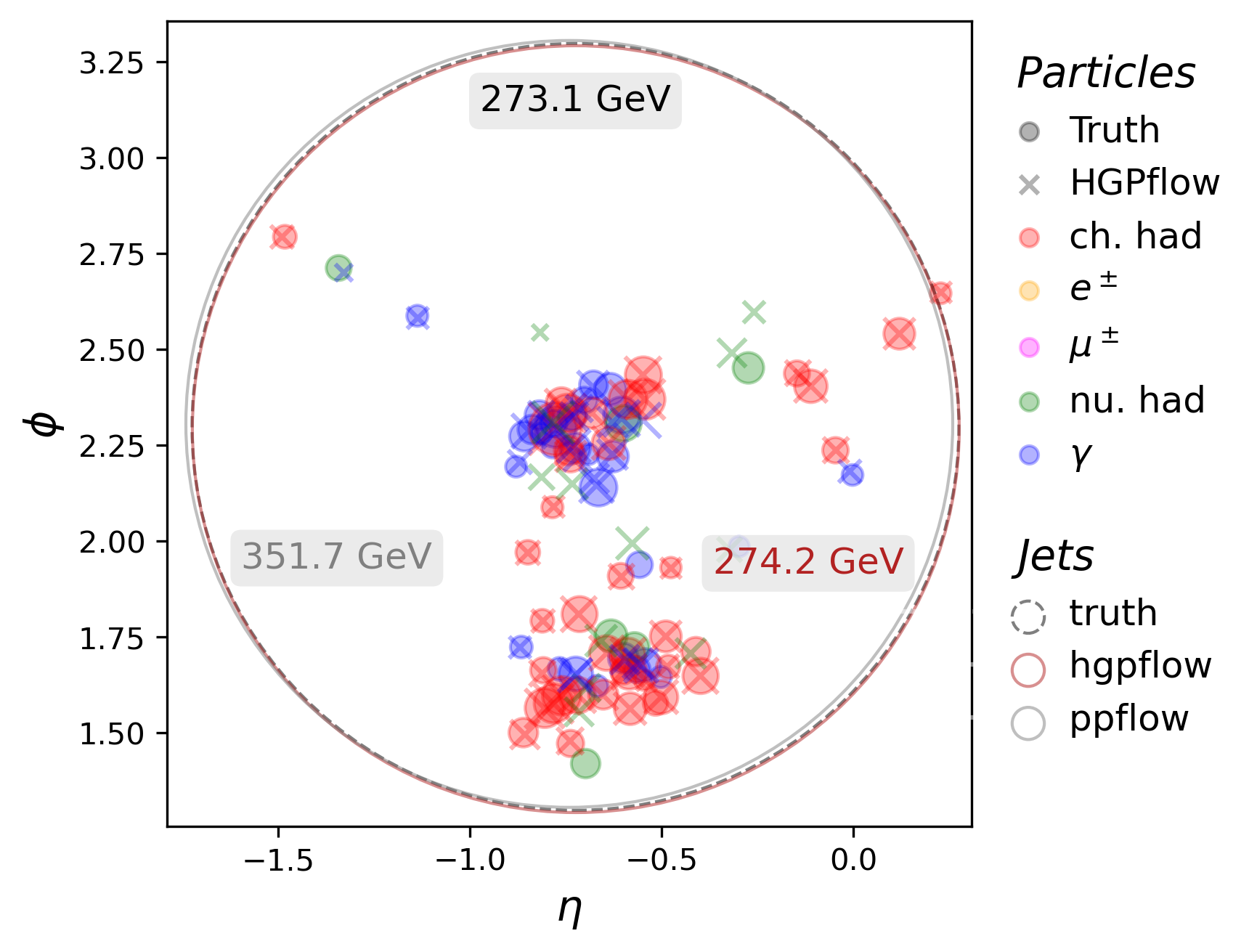}
        \caption{$Z\left( \nu \bar{\nu} \right) H \left( b \bar{b} \right)$}
        \label{fig:zhbb_lnd2}
    \end{subfigure}
    \caption{Examples of full events showing angular coordinates and classes of truth and predicted particles for three different processes. The marker size is proportional to the logarithm of particle \pt. Circles indicate the location and \pt values of leading jets formed from truth particles, HGPflow predictions, and PPflow objects using the anti-$k_\mathrm{T}$ algorithm~\cite{Cacciari:2008gp}. In (c) the jet radius parameter is $R=1.0$ whereas (a) and (b) have $R=0.4$.}
    \label{fig:event_displays}
\end{figure*}

\subsection{COCOA samples}\label{sec:samples}

Our main datasets are $pp$ collisions at $\sqrt{s}=13\ \mathrm{TeV}$ generated and showered with \textsc{Pythia8} \cite{Pythia8}. The generated events include initial and final state radiation and underlying event, but no pileup. The detector response is simulated using COCOA \cite{COCOA} -- a nearly hermetic calorimeter model based on \textsc{Geant4} \cite{geant1,geant2,geant3}. We use the default configuration of the COCOA calorimeter, comprising three electromagnetic calorimeter (ECAL) layers followed by three hadronic calorimeter (HCAL) layers.
In each layer, cells are defined by segmenting both $\eta$ and $\phi$ into $D$ divisions, where $D=[256,256,128]$ for the three ECAL layers and $D=[64,64,32]$ for the HCAL. 

Charged particle tracks are obtained by solving the helix equations in an axial 3.8T magnetic field, and material interactions are emulated through a parametric smearing applied to the $\eta$, $\phi$, and $q/p$ track parameters. Tracks are required to reach the calorimeter and to originate from a vertex with a transverse radius smaller than 75 mm and a longitudinal displacement less than 420 mm in order to be reconstructed. For our purposes, we remove all material from the tracking volume in the simulation, up until the iron solenoid layers preceding the calorimeter. Finally, a built-in topological clustering algorithm builds clusters (referred to as ``topoclusters'') by considering the energy-over-noise values of cells for three successive thresholds.

Three different physics hard scatter processes are used for our results: inclusive dijet, inclusive \ttbar, and boosted $Z(\nu\nu)H(b\overline{b})$ production. The number of events in each sample is listed in Tab.~\ref{tab:samples}. In the Higgs sample, the $Z$ boson is forced to decay to neutrinos, and a transverse momentum of the Higgs boson of at least 400~GeV is required. Single-pion samples are also simulated in COCOA to evaluate performance in a standard particle flow setup. The training of the ML algorithms is performed solely on the dijet sample while the other processes are for performance evaluation. Figure~\ref{fig:event_displays} shows one event display for each of the three physics processes simulated in COCOA.

\begin{table}[!h]
    \centering
\begin{tabular}{c|c|ccc}
\hline
\multirow{2}{*}{Detector} & \multirow{2}{*}{Process}                                   & \multicolumn{3}{c}{Statistics}          \\
                          &                                                            & train & val. & test                     \\ \hline
\multirow{4}{*}{COCOA}    & $p^+p^+ \rightarrow q\overline{q}$                         & 250k   & 10k   & 35k                      \\
                          & single $\pi^+$                                             & --    & --   & 30k / $p_\mathrm{T}$ bin \\
                          & $p^+p^+ \rightarrow t\overline{t}$                         & --    & --   & 20k                      \\
                          & $p^+p^+ \rightarrow Z(\nu\overline{\nu}) H(b\overline{b})$ & --    & --   & 10k                      \\ \hline
CLIC                      & $e^+e^- \rightarrow q\overline{q}$         & 1M    & 5k   & 20k                      \\ \hline
\end{tabular}
    \caption{Samples used for training and performance testing.}
    \label{tab:samples}
\end{table}

\begin{figure*}[ht!]
    \centering
    \begin{minipage}{0.335\textwidth}
        \hspace{-5mm}
        \vspace{1mm}
        \includegraphics[width=\textwidth]{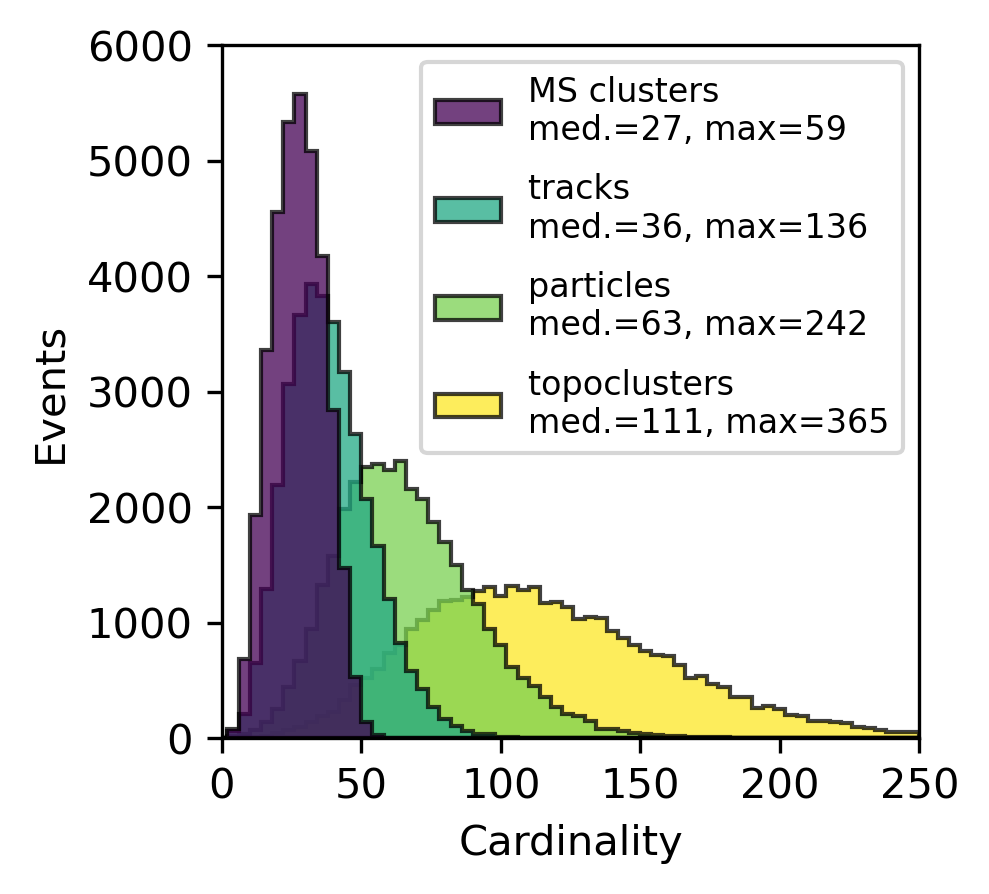}
        \subcaption{}
    \end{minipage}
    \begin{minipage}{0.31\textwidth}
        \hspace{-4mm}
        \includegraphics[width=\textwidth]{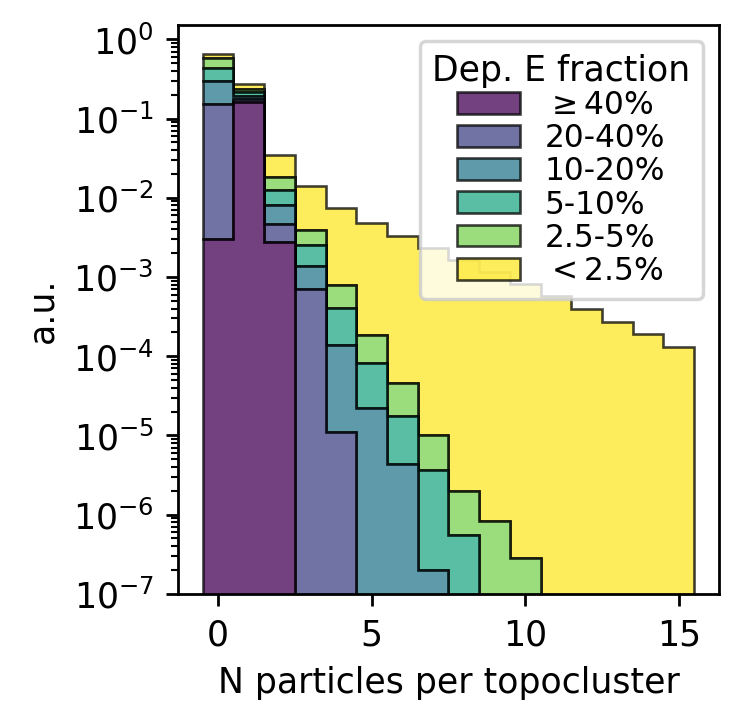}
        \subcaption{}
    \end{minipage}
    \begin{minipage}{0.31\textwidth}
        \hspace{-4mm}
        \includegraphics[width=\textwidth]{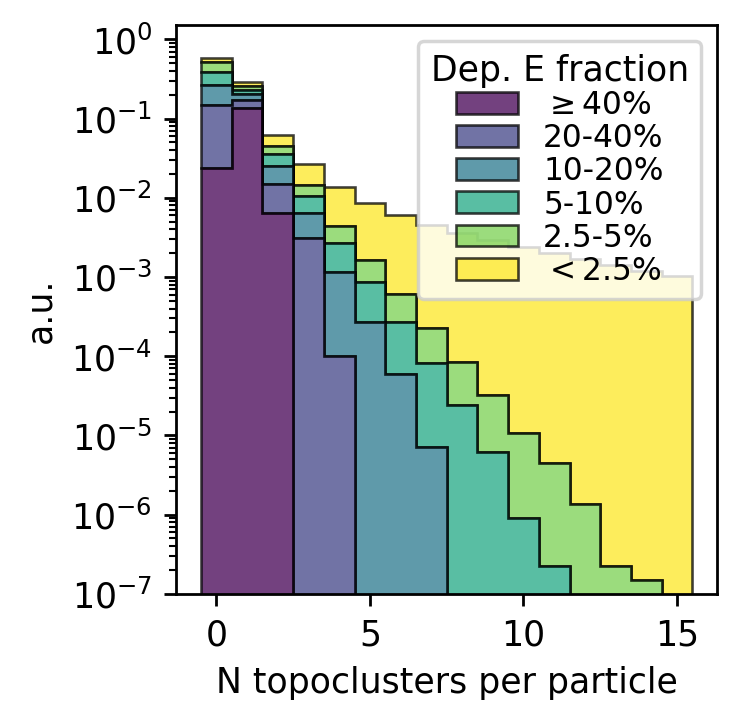}
        \subcaption{}
    \end{minipage}
    
    \caption{Cardinality and energy sharing characteristics in dijet events simulated in COCOA. (a) the number of objects in each event, where particles follow the target definition for COCOA. (b) and (c) show the number of particles associated with each topocluster and vice versa. The different histograms denote fractions of total deposited energy and are stacked.}
    \label{fig:cardinality-and-incidence}
\end{figure*}

Some key characteristics of the COCOA dijet events used to train HGPflow are shown in Fig.~\ref{fig:cardinality-and-incidence}. First, Fig.~\ref{fig:cardinality-and-incidence}a shows how many particles, tracks, and topoclusters are present in the events, as well as the number of partitions resulting from the MS clustering. The particles fulfill the requirements for target particles in COCOA defined in Tab.~\ref{tab:target}. 
Figures~\ref{fig:cardinality-and-incidence}b and \ref{fig:cardinality-and-incidence}c can be interpreted as the number of entries per row and per column of the incidence matrix, respectively. The entries are separated into different histograms based on the fraction of total deposited energy they carry. While it is true that most topoclusters have at least one particle accounting for over 40\% of its energy, long tails indicate that many additional particles are typically present, contributing smaller fractions of energy. The same is true for particles, with some particles dispersing their energy into as many as ten different topoclusters each containing $5-10\%$ of the energy.

\subsection{CLIC samples}

To study the possible extension of HGPflow to high-granularity calorimeters, we use a sample of simulated dijet events produced in $\sqrt{s}=380\ \mathrm{GeV}$ $e^+e^-$ collisions at the Compact Linear Collider (CLIC). The sample is publicly available at \cite{pata_zenodo} where it is part of the larger training sample used to train MLPF in \cite{Pata:2023rhh}. The samples are produced using \textsc{Pythia8} and the CLIC detector simulation \cite{CLICdp:2017vju, CLICdp:2018vnx}, which is based on \textsc{Geant4} and the \textsc{Key4HEP} software \cite{Ganis:2021vgv}. 

Compared to $pp$ collisions at the LHC, $e^+e^-$ collisions at CLIC are cleaner, having neither pileup nor underlying event. The calorimeter of the CLIC detector is also highly granular, comprising 40 ECAL layers with a cell pitch of $5\times 5 \ \mathrm{mm}^2$ and 60 HCAL layers with a $30\times 30 \ \mathrm{mm}^2$ cell pitch. The calorimeters are surrounded by a solenoid magnet providing a field strength of 4T. Unlike COCOA, the CLIC detector simulation includes tracking material and realistic track reconstruction. Each event contains particle flow objects reconstructed by the \pd algorithm \cite{Marshall:2015rfa}. Calorimeter clusters are provided with a predefined calibration applied to their energy.


\section{Target definition and training}
\label{sec:target}
\subsection{Fiducial particle definition}\label{sec:fiducial}

Defining which particles comprise the target is a key design choice when approaching particle reconstruction as a DL task. In actual detectors, a significant amount of material lies upstream of the calorimeters in the tracking volume. Interactions with this material can cause non-trivial alterations to the set of stable particles leaving the interaction, particularly in the cases of photon conversions and electron bremsstrahlung. A fiducial electron definition in the form of angular criteria is necessary to decide when radiated energy should be merged with an electron and when it should be considered as a separate photon in the target set. Similarly, depending on the angular separation of two conversion electrons entering the calorimeter, a fiducial photon definition would distinguish cases where the resulting clusters should be considered merged or resolved, relative to detector resolution.

Given the detector dependence of designing fiducial criteria for electrons and photons, we have decided to avoid a full treatment in this paper, focusing instead on the feasibility of reconstructing full events. In COCOA, our choice is easily realized by the removal of tracking material, such that upstream interactions only occur in the iron layer immediately preceding the calorimeter. For CLIC, the situation is more nuanced, since the presence of tracking material causes significant rates of particles to be created upstream of the calorimeter. These particles have a generator status code of 0, easily distinguishing them from stable particles (status code 1) and unstable particles (status code 2) provided by \textsc{Pythia}. When an interaction marks the end of a neutral parent particle trajectory (for instance, in photon conversion), it will have no associated energy deposit in the detector. It is problematic to train a physically-motivated model such as HGPflow to reconstruct particles that do not leave any trace in the detector. On the other hand, it is incorrect to simply remove such particles from the ground truth.

To resolve this conflict, we propose a practical compromise that separates the definition of the ground truth and the definition of the target used in the training objective, as defined in Tab.~\ref{tab:target}. This separation ensures that performance evaluation is based on a clear standard while allowing the freedom to design a training target suitable for each DL model. 

\begin{table}[!ht]
\begin{tabular}{c|cc|cc}
\hline
                & \multicolumn{2}{c|}{COCOA}                       & \multicolumn{2}{c}{CLIC}                              \\
              & Truth                  & Target                 & Truth                     & Target                    \\ \hline
$\nu$ veto              & $\checkmark$           & $\checkmark$           & $\checkmark$              & $\checkmark$              \\
$p_\mathrm{T}\ (E)$ [GeV]              & > 1 & > 1 & (> 0.01) & (> 0.10) \\
$|\eta|$                & \textless{}3           & \textless{}3           & \textless{}4              & \textless{}4              \\
Generator status        & 1                      & 1                      & 1                         & 0, 1, 2                   \\
Interacts with detector & $-$                    & $\checkmark$           & $-$                       & $\checkmark$          \\
\hline
\end{tabular}
\caption{Definitions of the ground truth and the training target for particles in the COCOA and CLIC datasets. Particles that interact with the detector are defined as those that are associated with a track or an energy deposit in the calorimeter.}
\label{tab:target}
\end{table}

The target definition we use to train MLPF is identical to the one in \cite{Pata:2023rhh} except that we also include interacting particles with generator status 0 and 2, as this significantly improved alignment with our ground truth definition. Unlike \hg, the training objective in MLPF requires each particle to be associated with a unique track or calorimeter cluster. At low-\pt, neutral particles frequently fail to produce a leading energy contribution to any calorimeter cluster, making such an association ambiguous. For the COCOA dijet events we use for training, roughly 20\% of photons and 10\% of neutral hadrons fall into this category. In MLPF, such particles are merged kinematically with the nearest suitable target particle so that their energy can be accounted for in aggregate. Charged particles that absorb nearby neutral particles by this definition can thus have target properties that are modified with respect to their associated track and ground truth particle.
\subsection{Training}\label{sec:training}

The hyperparameters for the \hg model and training are shown in Tab.~\ref{tab:hyperparam}. \hg is written using the \textsc{PyTorch} library~\cite{pytorch}. The learning rate is scheduled using a single cosine decay cycle \cite{loshchilov2017sgdrstochasticgradientdescent} with warmup. No hyperparameter optimization was performed for \hg, while for MLPF we used the model with hyperparameters obtained from the optimization performed in \cite{Pata:2023rhh}. We use version 1.6.2 of the latest public MLPF code~\cite{pata_zenodo_code}. The first stage of \hg was trained for roughly 57 (100) hours on the COCOA (CLIC) sample, corresponding to 100 epochs on a single NVIDIA RTX A6000 graphics processing unit (GPU). The second stage of training progressed at roughly 10 (5) epochs per hour for 30 (18) epochs. During inference, the threshold for the indicator prediction in \hg is set to 0.5 (0.65) for COCOA (CLIC).


\begin{table}[!ht]
\centering
\begin{tabular}{l|r}
\hline
\; Optimizer & AdamW \cite{Loshchilov2017DecoupledWD} \\
\; Learning rate (max) & 0.001 \\
\hline
\; Batch size - & \\
\; \; \; \;stage one & 256 \\
\; \; \; \;stage two & 512 \\
\hline
\; Number of epochs - & \\
\; \; \; \;stage 1 & 100 \\
\; \; \; \;stage 2 & 30 \\
\hline
\; Maximum number of hyperedges ($K$) & 60 \\
\; Embedded dimension & 128 \\
\; Number of iterative refinements & 12 \\
\; Number of trainable parameters & 2.1M \\
\hline
\end{tabular}
\caption{Hyperparameters used for training HGPflow on the COCOA dataset.}
\label{tab:hyperparam}
\end{table}

\vspace{10mm}
\section{Results}\label{sec:results}
We evaluate performance using particle-level, jet-level, and event-level metrics on the samples defined in Sec.~\ref{sec:samples}. For each event, the sets of predicted particles from each partition are pooled together and compared with the set of truth particles in the full event. The predictions from the other reference algorithms -- \pf, \pd, or MLPF -- are passed through the same performance pipeline as \hg. Figure~\ref{fig:event_displays} shows truth particles and particles from \hg for example dijet, $t\overline{t}$, and boosted $Z(\nu\nu)H(b\overline{b})$ events.

\begin{figure*}[ht!]
\begin{subfigure}{.33\textwidth}
  \centering
  \includegraphics[width=\linewidth]{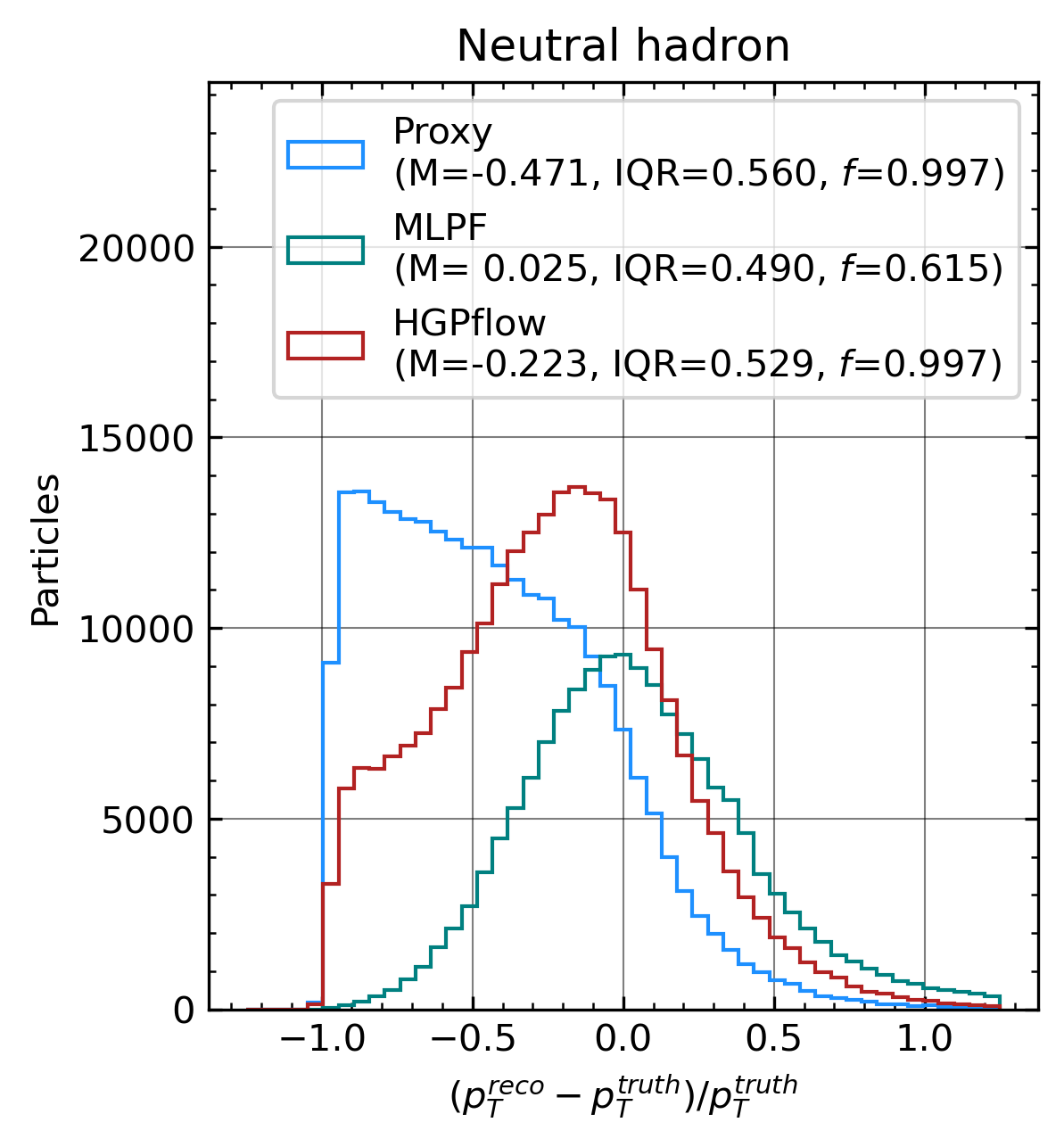}
  \caption{}
  \label{fig:cocoa_part_res_nh_pt}
\end{subfigure}%
\begin{subfigure}{.33\textwidth}
  \centering
  \includegraphics[width=\linewidth]{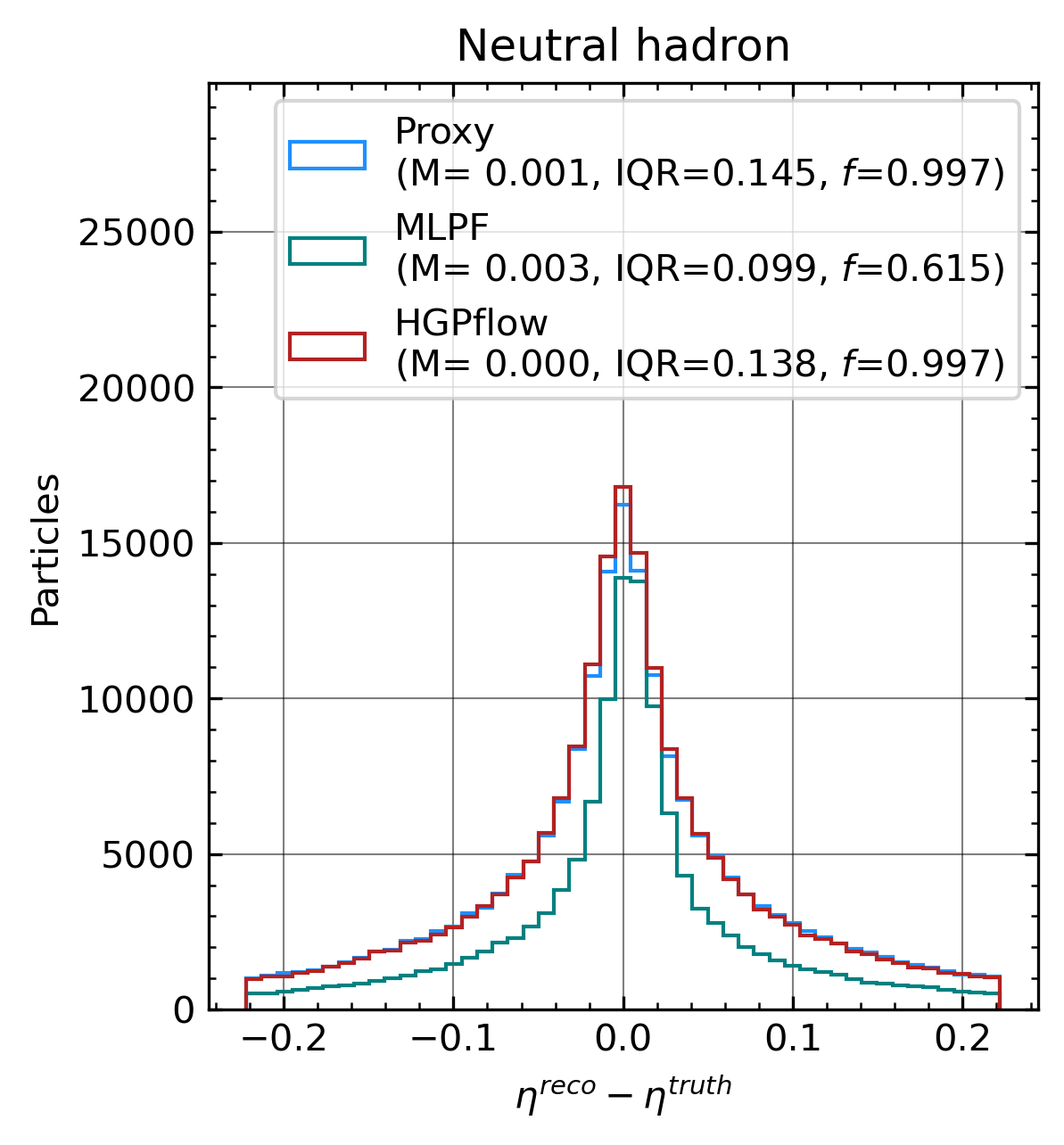}
  \caption{}
  \label{fig:cocoa_part_res_nh_eta}
\end{subfigure}%
\begin{subfigure}{.33\textwidth}
  \centering
  \includegraphics[width=\linewidth]{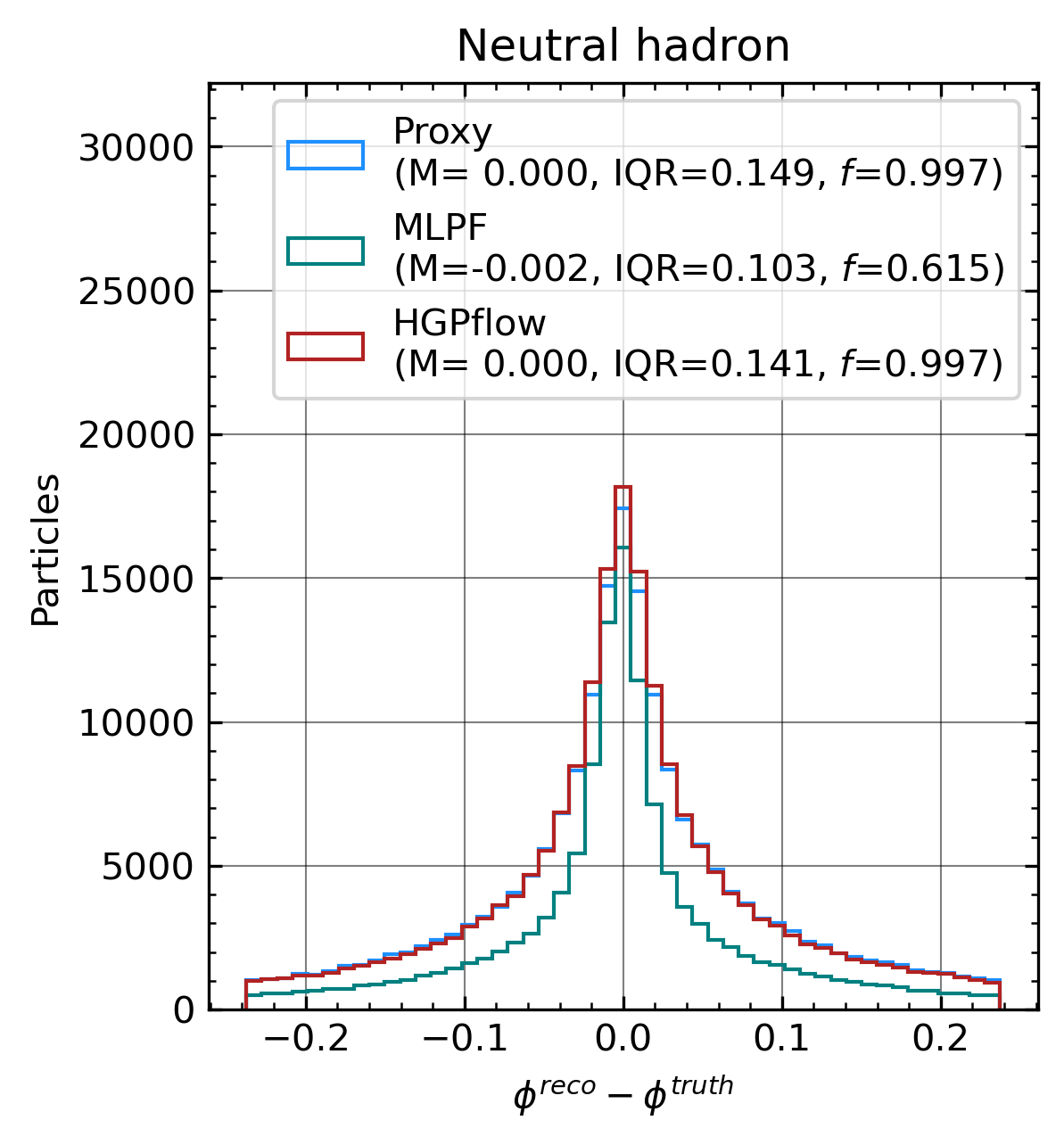}
  \caption{}
  \label{fig:cocoa_part_res_nh_phi}
\end{subfigure}%

\begin{subfigure}{.33\textwidth}
  \centering
  \includegraphics[width=\linewidth]{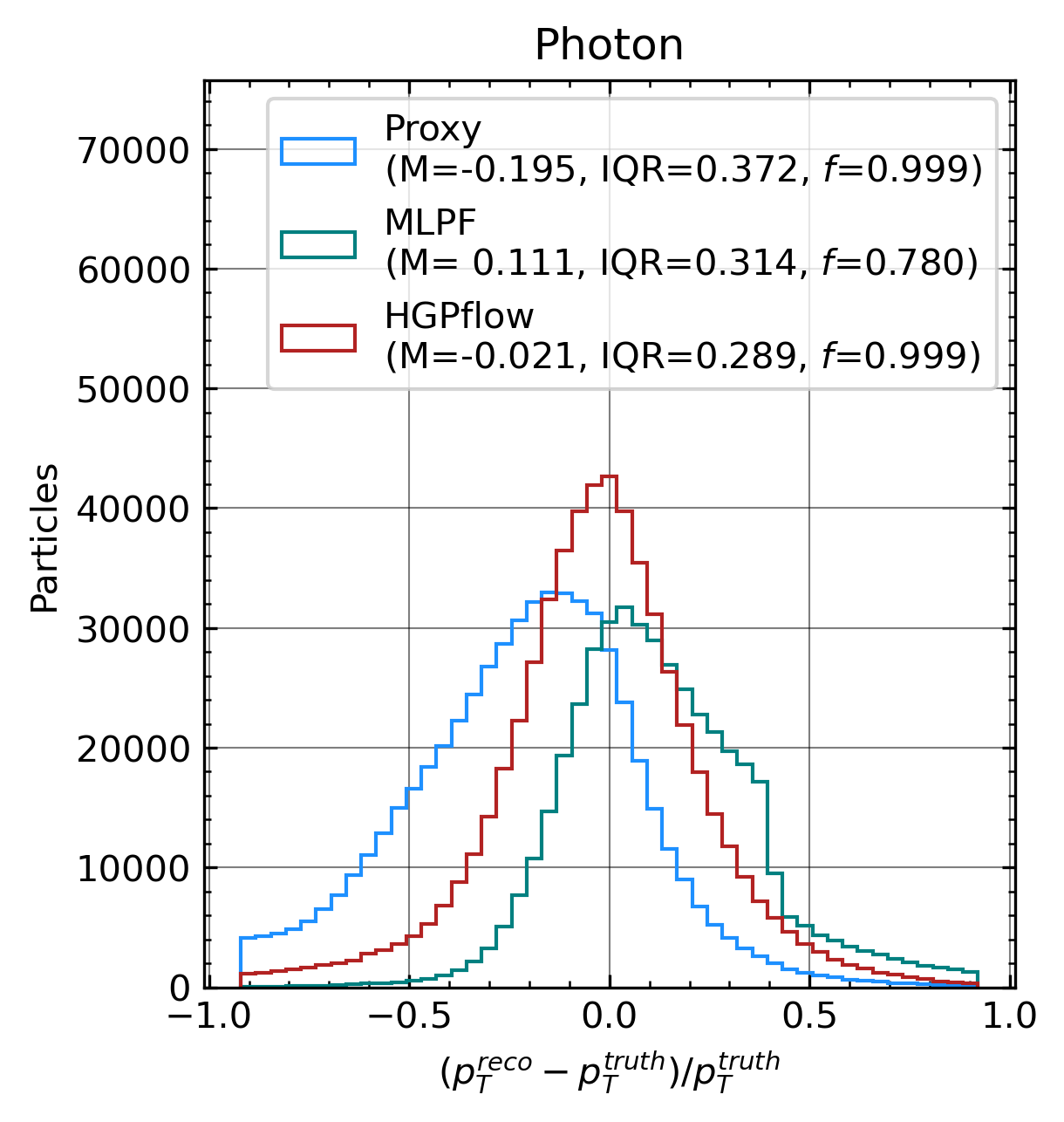}
  \caption{}
  \label{fig:cocoa_part_res_ph_pt}
\end{subfigure}%
\begin{subfigure}{.33\textwidth}
  \centering
  \includegraphics[width=\linewidth]{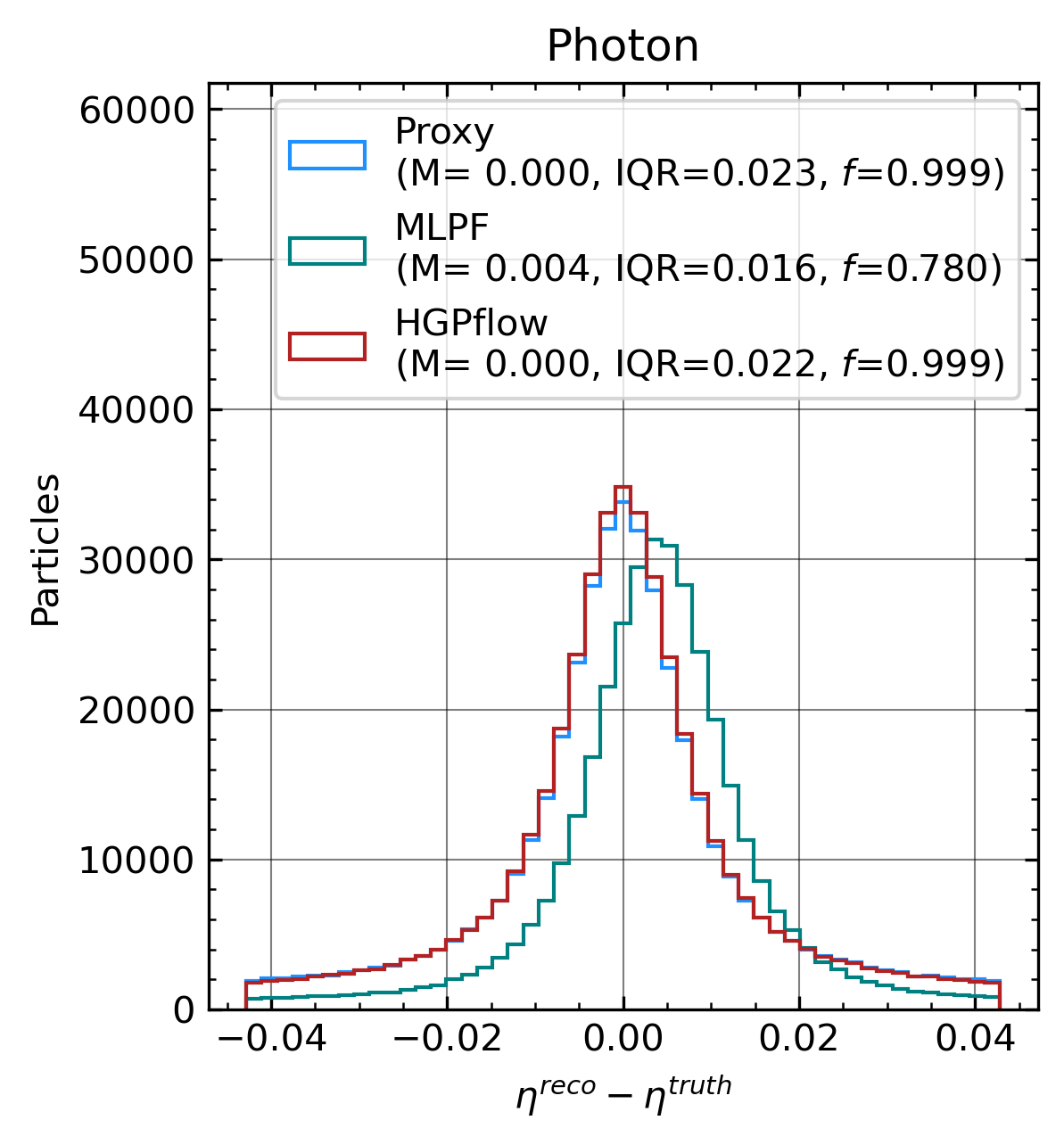}
  \caption{}
  \label{fig:cocoa_part_res_ph_eta}
\end{subfigure}%
\begin{subfigure}{.33\textwidth}
  \centering
  \includegraphics[width=\linewidth]{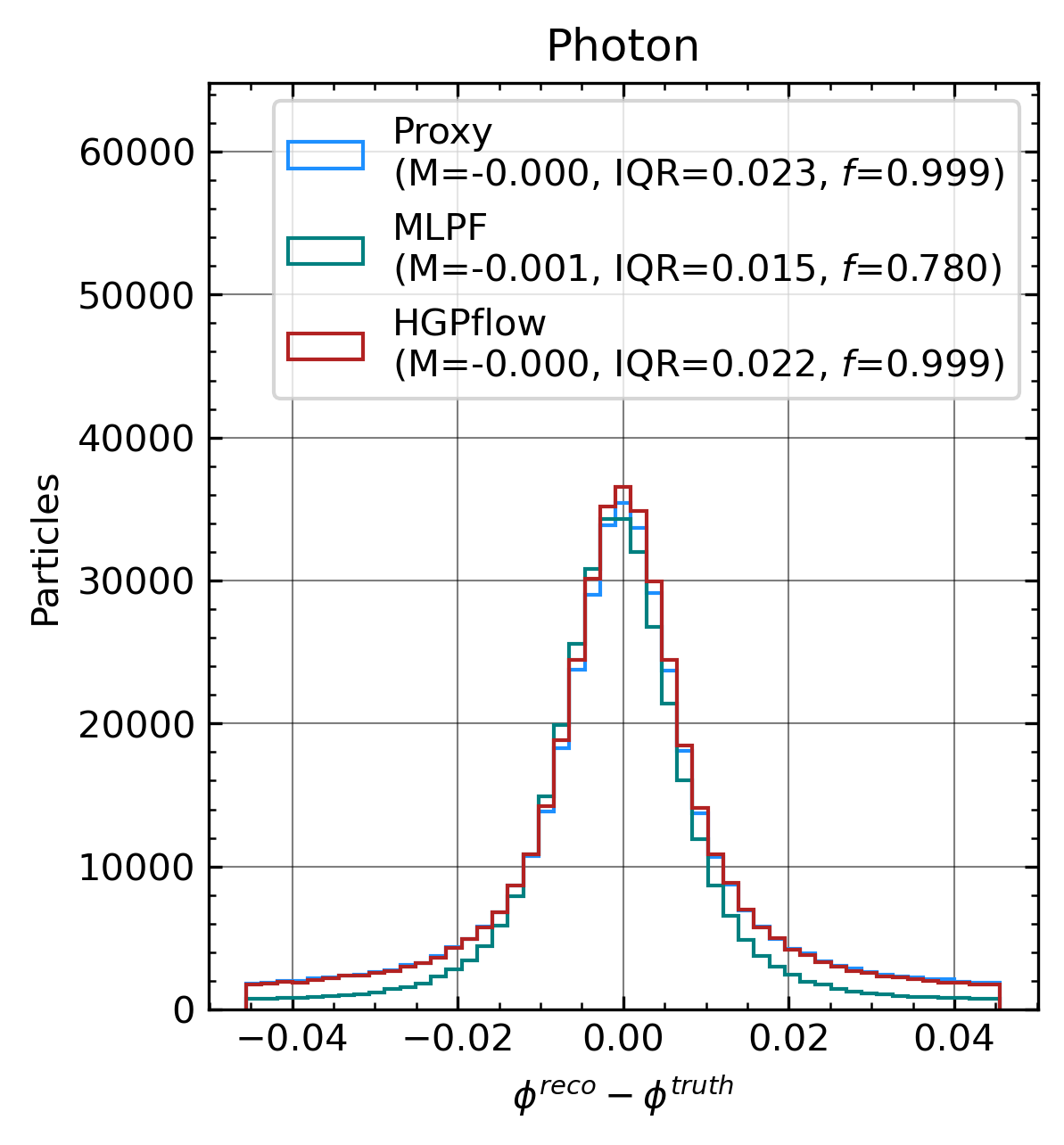}
  \caption{}
  \label{fig:cocoa_part_res_ph_phi}
\end{subfigure}
\caption{Residual distributions for \pt, $\eta$, and $\phi$ of the neutral hadrons (a, b, c) and photons (d, e, f) matched to reconstructed particles from \hg and MLPF. The ``Proxy'' distribution refers to particles formed using only the incidence matrix and class predictions of \hg without applying any learned correction. The metrics M and IQR refer to the median and interquartile range, respectively, while $f$ is the efficiency of matching truth particles.}
\label{fig:neutral_performance}
\end{figure*}

\subsection{Dijet sample (COCOA)} \label{sec:dijet_cocoa}

Since the \hg algorithm was trained exclusively on dijet events, we first investigate its performance on an independent test set of dijet events. To study particle-level performance, we use the Hungarian algorithm \cite{kuhn1955hungarian} to create pairs between the sets of predicted and truth particles. The matching is performed separately for charged and neutral particles using the following cost function:
\begin{equation}
  c_\mathrm{Hung.} = \sqrt{(\Delta p_\mathrm{T} / p_\mathrm{T}^\mathrm{targ})^2 + \Delta \eta^2 + \Delta \phi^2}
\end{equation}
In case the two sets have unequal sizes, a subset of particles from the larger set will be unpaired. For neutral particles, such cases of lower or higher cardinality with respect to the set of true particles are measured in terms of efficiency and fake rate, as follows:
\begin{align}
    &\epsilon = N(\mathrm{matched})/N(\mathrm{true}) \label{eq:efficiency} \\
    &f = N(\mathrm{unmatched})/N(\mathrm{pred.}) \label{eq:fakerate}
\end{align}
The cardinality of charged particles, on the other hand, is determined in both \hg and MLPF by the number of tracks in the event.

Figure~\ref{fig:neutral_performance} shows reconstruction quality in terms of neutral particle properties. The histograms only include reconstructed neutral particles that are paired with truth neutral particles, so the normalization of the histograms reflects the matching efficiency. It is also important to note that the histograms are dominated by low-\pt particles, due to the steeply falling spectrum. Comparing the \hg and proxy (Eq.~\ref{eq:proxy1}-~\ref{eq:proxy2}) histograms reveals the improvement obtained through the second stage of training. For both photons and neutral hadrons, the proxy quantities shown in blue already provide reasonable angular coordinates. For transverse momentum, on the other hand, the importance of a learned correction to the uncalibrated measurement from the calorimeter is clear. This is especially noticeable for neutral hadrons in Fig~\ref{fig:neutral_performance}a because of their intrinsically low energy response and reduced topoclustering efficiency, which skew their reconstructed \pt toward zero. Compared to MLPF, HGPflow has a higher efficiency in matching true neutral particles, but it exhibits an overall worse resolution. This apparent trade-off can be understood as arising from the different target definitions used by each algorithm, as discussed in Sec~\ref{sec:fiducial}. While MLPF focuses on the subset of neutral particles that dominate one or more topoclusters, HGPflow also targets the more challenging neutral particles that contribute only sub-dominant energy fractions.

\begin{figure}[!ht]
    \centering
    \includegraphics[width=0.50\textwidth]{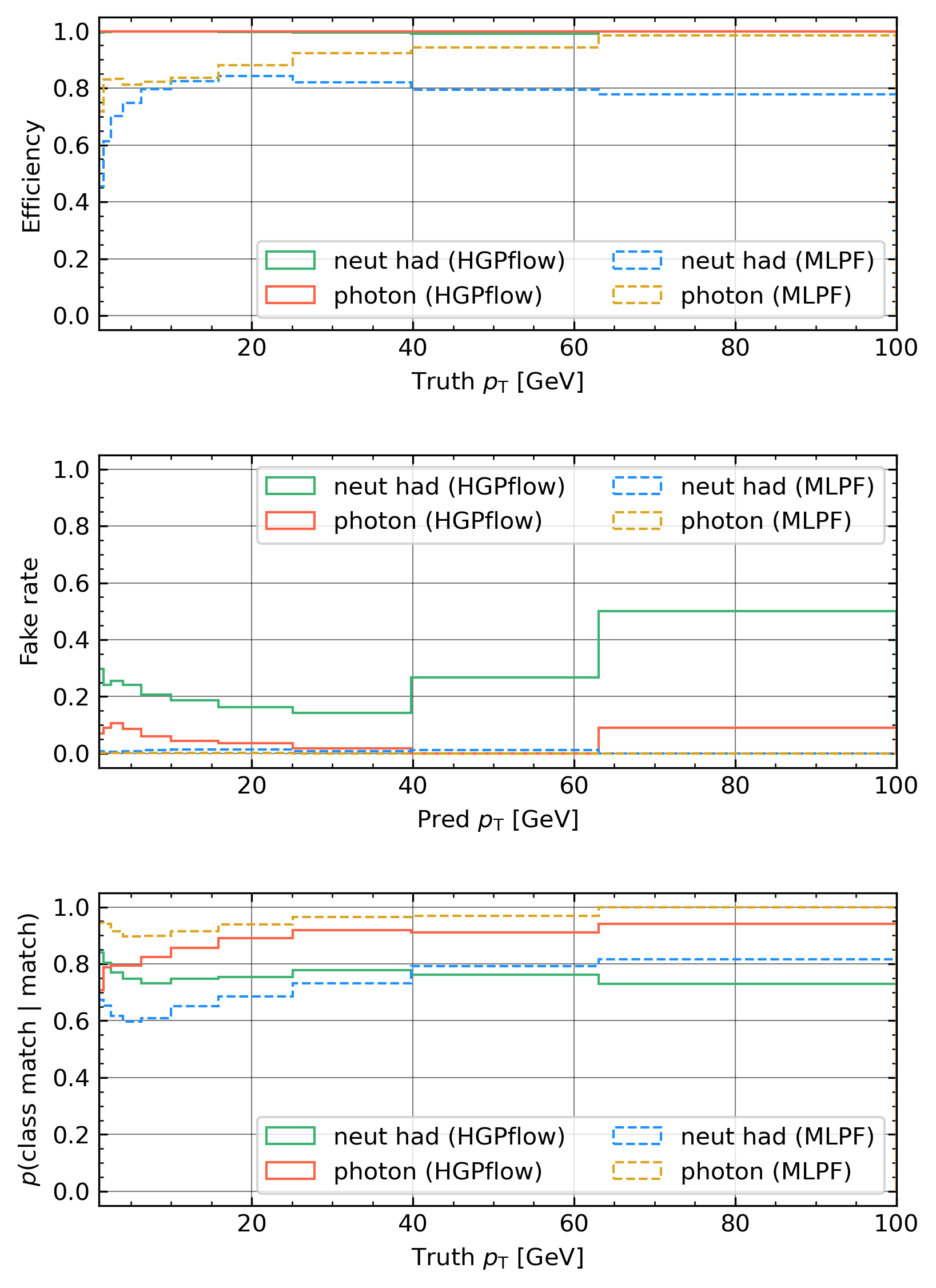}
    \caption{Top and middle panels show neutral particle efficiency and fake rate for HGPflow and MLPF, as defined in Eqs.~\ref{eq:efficiency}-\ref{eq:fakerate}. The bottom panel displays the fraction of pairs for which the reconstructed particle class matches that of its corresponding truth particle.}
    \label{fig:eff_fr_purity}
\end{figure}

The top panel of Fig.~\ref{fig:eff_fr_purity} shows the matching efficiency (Eq.~\ref{eq:efficiency}) for photons and neutral hadrons in bins of truth \pt. On the flip side, the middle of Fig.~\ref{fig:eff_fr_purity} shows the fake rate, i.e., the rate of unmatched particles predicted, as defined in Eq.~\ref{eq:fakerate}. Finally, considering predicted particles matched to truth photons or neutral hadrons, the bottom panel in Fig.~\ref{fig:eff_fr_purity} reports the fraction of correct class prediction among the pairs. Overall, \hg reconstructs neutral particles with nearly 100\% efficiency. On the other hand, MLPF has lower efficiency but a fake rate of nearly zero. As with Fig.~\ref{fig:neutral_performance}, here it is important to note the interplay between these metrics and the cardinality prediction. For \hg, tightening the indicator threshold during inference would lower the fake rate at some expense to efficiency. However, for MLPF, we suspect that the efficiency is limited by the merging procedure described in Sec~\ref{sec:fiducial}, wherein some neutral particles are absorbed into neighboring particles when creating the training target. Both algorithms perform comparably well at distinguishing photons and neutral hadrons -- a task that becomes considerably more difficult for low-\pt neutral particles.

Next, we evaluate the performance of \hg at the jet level. We form jets out of truth particles and out of predicted objects using the anti-$k_\mathrm{T}$ algorithm \cite{Cacciari:2008gp} implemented in \textsc{FastJet} \cite{Cacciari:2011ma} with $R=0.4$. In each event, truth jets are matched to reconstructed jets based on $\Delta R$ up to a maximal separation of 0.1.

\begin{figure*}[!ht]
\begin{subfigure}{.33\textwidth}
  \centering
  \includegraphics[width=\linewidth]{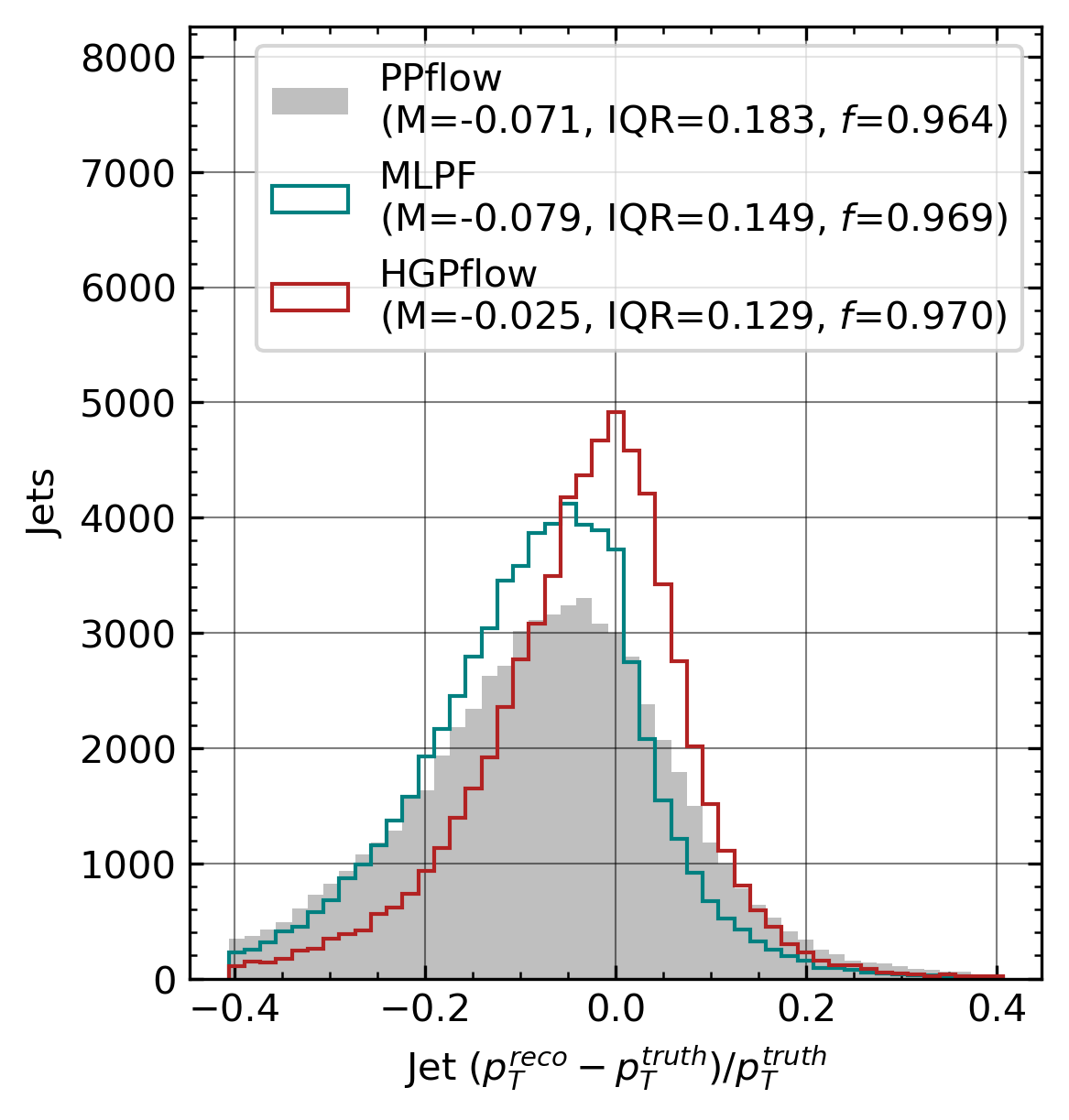}
  \caption{}
  \label{fig:cocoa_jet_res_pt}
\end{subfigure}%
\begin{subfigure}{.33\textwidth}
  \centering
  \includegraphics[width=\linewidth]{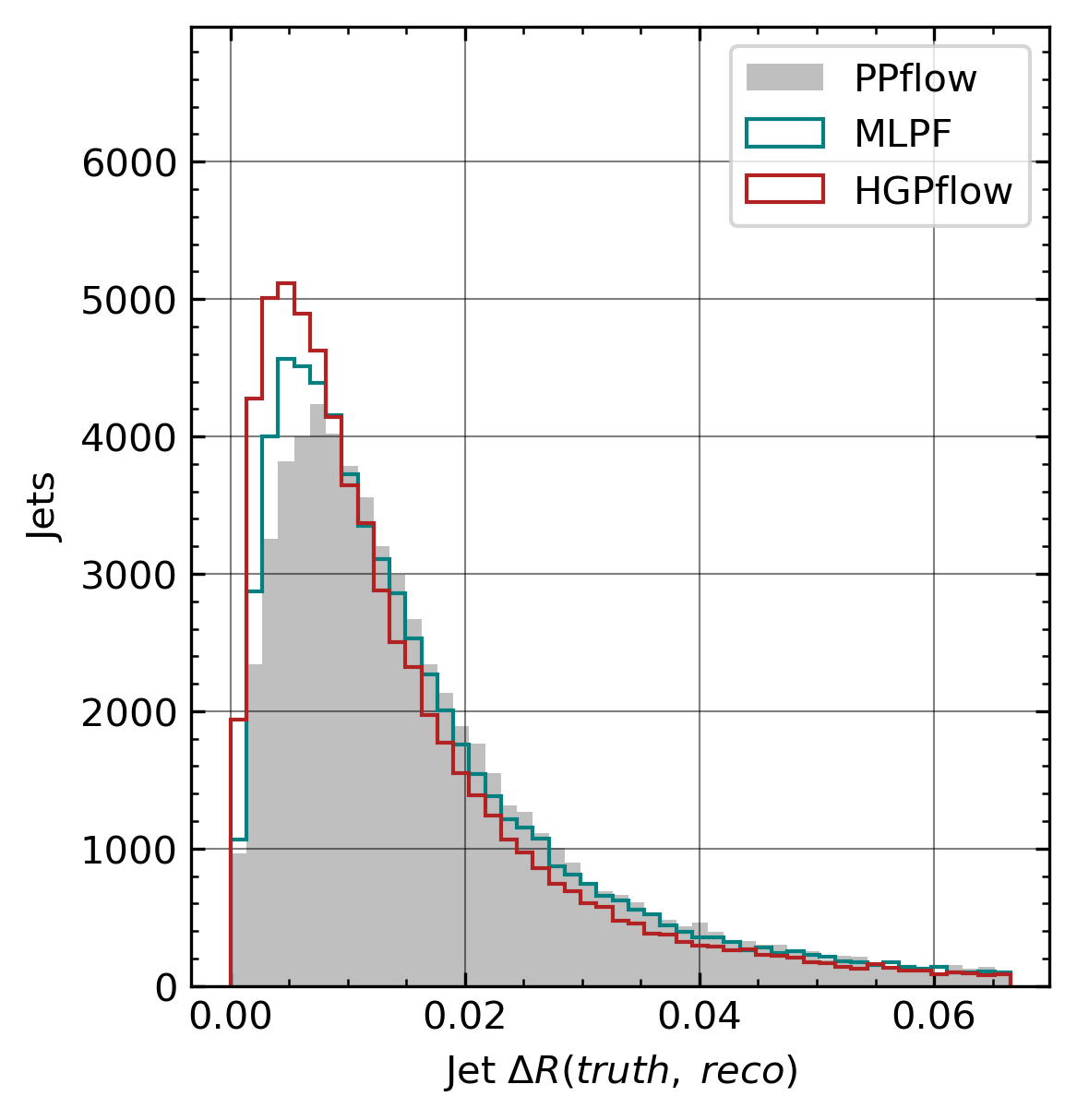}
  \caption{}
  \label{fig:cocoa_jet_res_dr}
\end{subfigure}%
\begin{subfigure}{.33\textwidth}
  \centering
  \includegraphics[width=\linewidth]{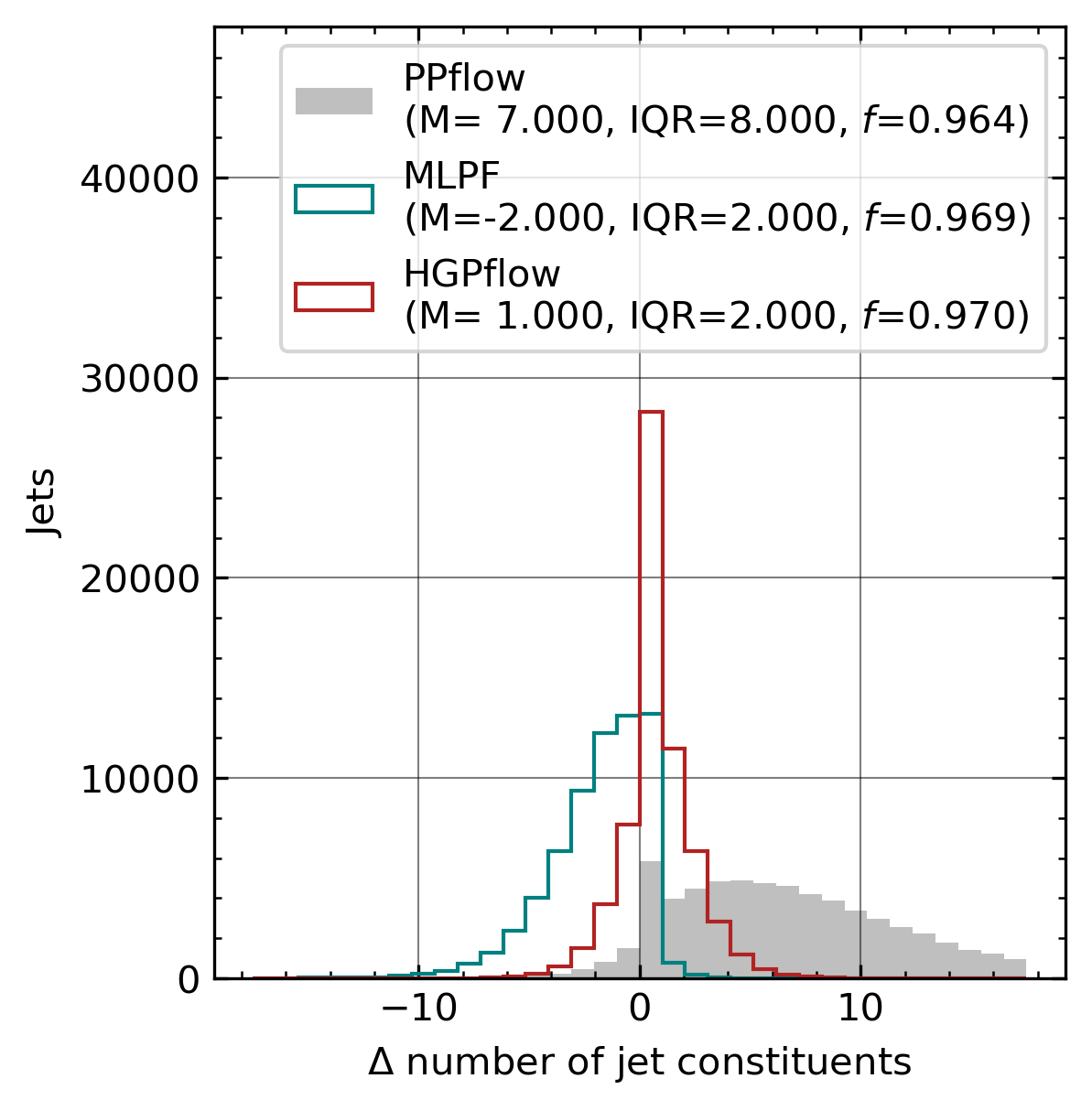}
  \caption{}
  \label{fig:cocoa_jet_res_nconst}
\end{subfigure}%
\caption{Matching characteristics between truth and reconstructed jets in terms of (a) jet \pt, (b) angular separation $\Delta R$, and (c) number of constituents $N_\mathrm{const.}$ for dijet events simulated in the COCOA detector. Up to two leading truth jets with \pt~$> 10$~GeV are taken from each event and matched to reconstructed jets based on $\Delta R$ up to a maximum of 0.1. The metrics M and IQR refer to the median and interquartile range, respectively, while $f$ is the efficiency of matching truth jets.}
\label{fig:leading_jet_residuals}
\end{figure*}

The distributions of relative jet \pt residuals are shown in Fig.~\ref{fig:leading_jet_residuals}. Both DL-based algorithms outperform the parameterized benchmark algorithm \pf, while \hg yields a narrower and less shifted \pt residual distribution compared to MLPF. 
Figure~\ref{fig:cocoa_jet_res_dr} also shows that \hg performs the best in terms of the angular separation between the predicted and truth jet. Finally, \hg models the number of jet constituents significantly better than \pf and MLPF, shown in Fig.~\ref{fig:cocoa_jet_res_nconst}. This can be understood in light of how cardinality is handled in each reconstruction approach. By associating the input and output sets through the incidence matrix, HGPflow is equipped to learn an arbitrary mapping between topoclusters and particles. In \pf, both topoclusters and tracks function as jet constituents, leading naturally to a cardinality larger than that of truth jets. This double counting can be suppressed in MLPF by assigning a subset of the topoclusters to the reject class. The underestimation observed for MLPF likely stems from its target definition, which is not designed for cases of multiple contributions to a given topocluster and merges neutral particles that are not leading in any topocluster.

\begin{figure*}[ht!]
\centering
\includegraphics[width=1.0\textwidth]{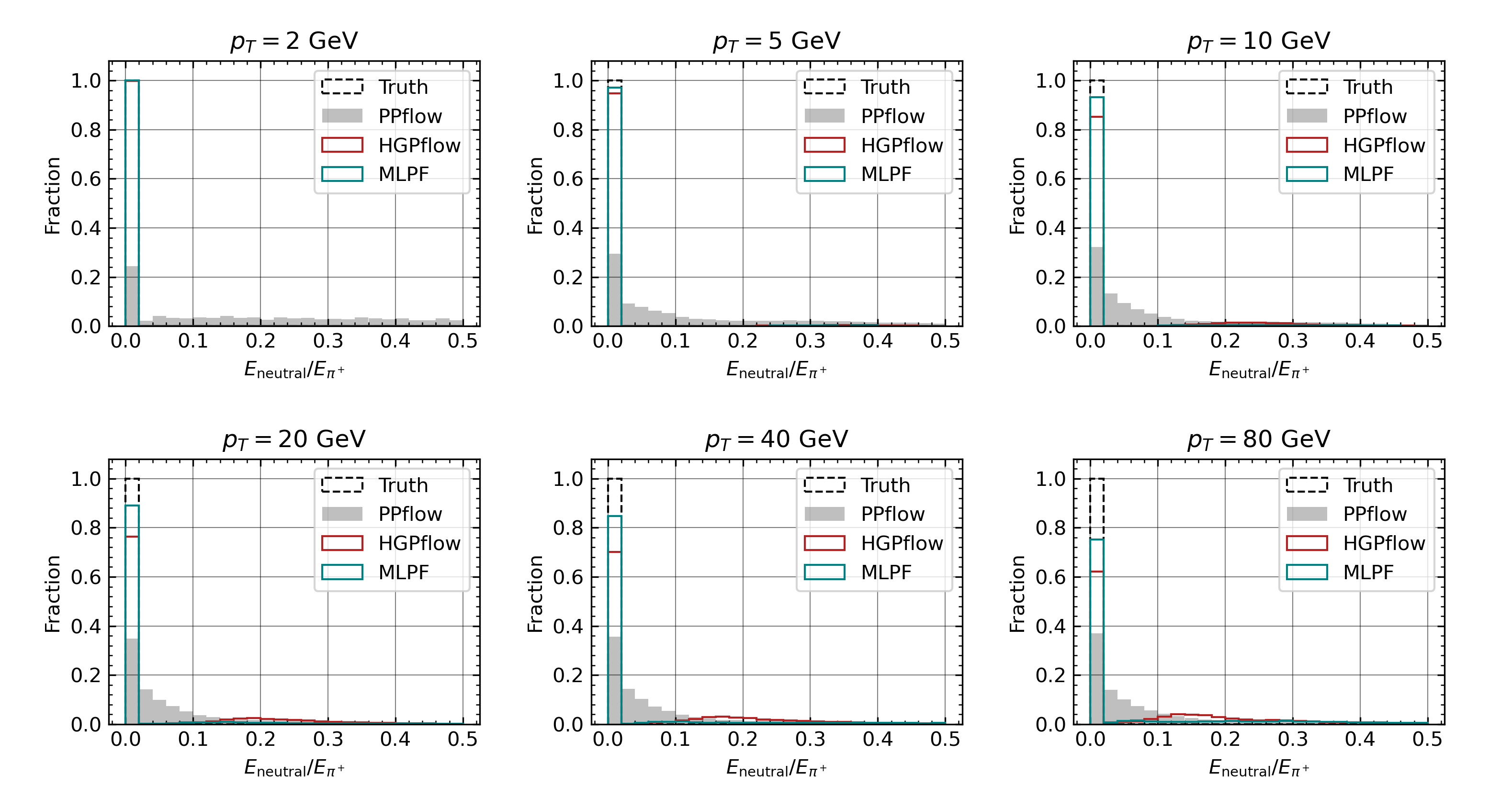}
\caption{The ratio of predicted neutral energy ($E_\mathrm{neutral}$) relative to charged energy $(E_{\pi^+})$ in events containing a single truth $\pi^+$. The subfigure title indicates the transverse momentum of the charged pion in each case. Each distribution is normalized, and the closer to the truth, the better.}
\label{fig:pion_residuals}
\end{figure*}

\subsection{Charged pion sample (COCOA)}

The traditional approach to particle flow reconstruction, represented by \pf, has at its core a model of calorimeter response for charged pions. Therefore, we compare the performance of the DL algorithms with that of \pf on samples of single $\pi^+$ generated uniformly in $|\eta|<3$ with fixed \pt values ranging from 2~GeV to 80~GeV. These are out-of-distribution examples, as the DL algorithms were not trained on isolated pions, and fewer than 1\% of charged pions in the training dataset have \pt~$>20$~GeV. In these cases, the key metric is how efficiently the algorithm predicts a single charged pion without mistaking calorimeter energy fluctuations for additional neutral particles. In practice, this capability allows a particle flow algorithm to incorporate tracks without double-counting.

Figure~\ref{fig:pion_residuals} shows the fraction of single-$\pi^+$ examples for which additional neutral particles are predicted, binned in the ratio of total neutral energy to true $\pi^+$ energy. Both \hg and MLPF outperform the \pf baseline, yielding significantly larger fractions in the first bin (0-2\%). For pions below 10~GeV, \hg predicts neutral energy levels below 2\% of the pion energy in over 85\% of the cases. The task becomes more challenging with increasing \pt, with \hg predicting the same $\leq 2\%$ neutral energy fraction for 62\% of pions with \pt~$=80$~GeV. Compared to MLPF, \hg predicts larger fractions of neutral particle energy for all the \pt bins. This can be understood as the same effect observed in Sec.~\ref{sec:dijet_cocoa}, where \hg was characterized by higher efficiency and higher fake rate than MLPF in Fig.~\ref{fig:eff_fr_purity}. 

The neutral particle energy predicted by HGPflow exhibits a broad peak between 10\% and 30\% of the $\pi^+$ energy. In contrast to \pf, where any fraction of a calorimeter cluster can be associated with neutral particles, \hg is trained to mimic energy assignments seen during training, leading to more discrete predictions. We observed that in over 90\% of the single pion examples, \hg predicts two or fewer neutral particles with energy fractions above 2\%. Overall, Fig.~\ref{fig:pion_residuals} demonstrates that both DL algorithms successfully extrapolate to an energy association task typically used to tune parametric particle flow models.

\begin{figure*}[!ht]
    \centering
    \begin{subfigure}[c]{0.33\textwidth}
        \includegraphics[width=\textwidth]{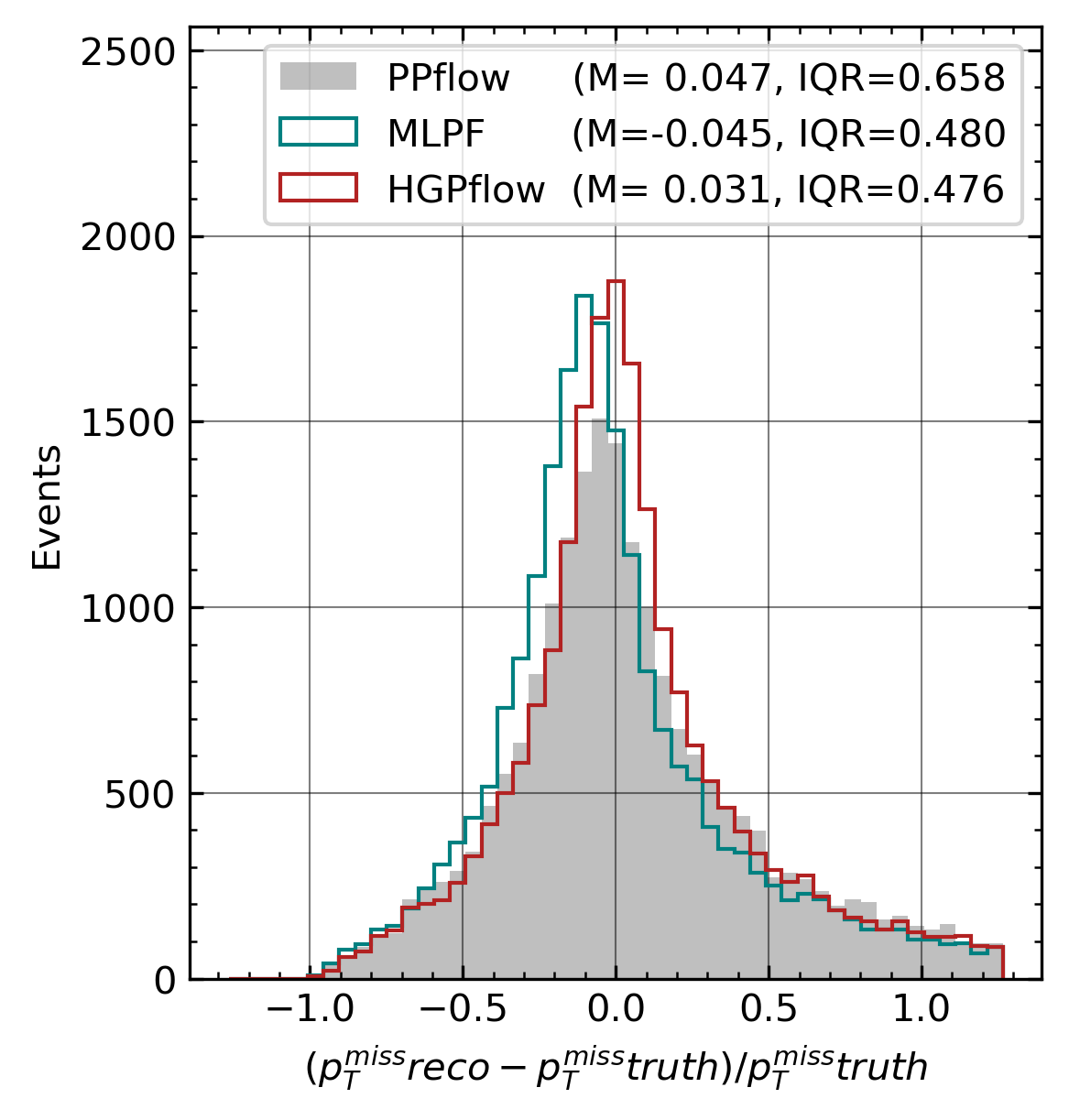}
        \caption{}
        \label{fig:ttbar_met_residual}
    \end{subfigure}
    \begin{subfigure}[c]{0.33\textwidth}
        \includegraphics[width=\textwidth]{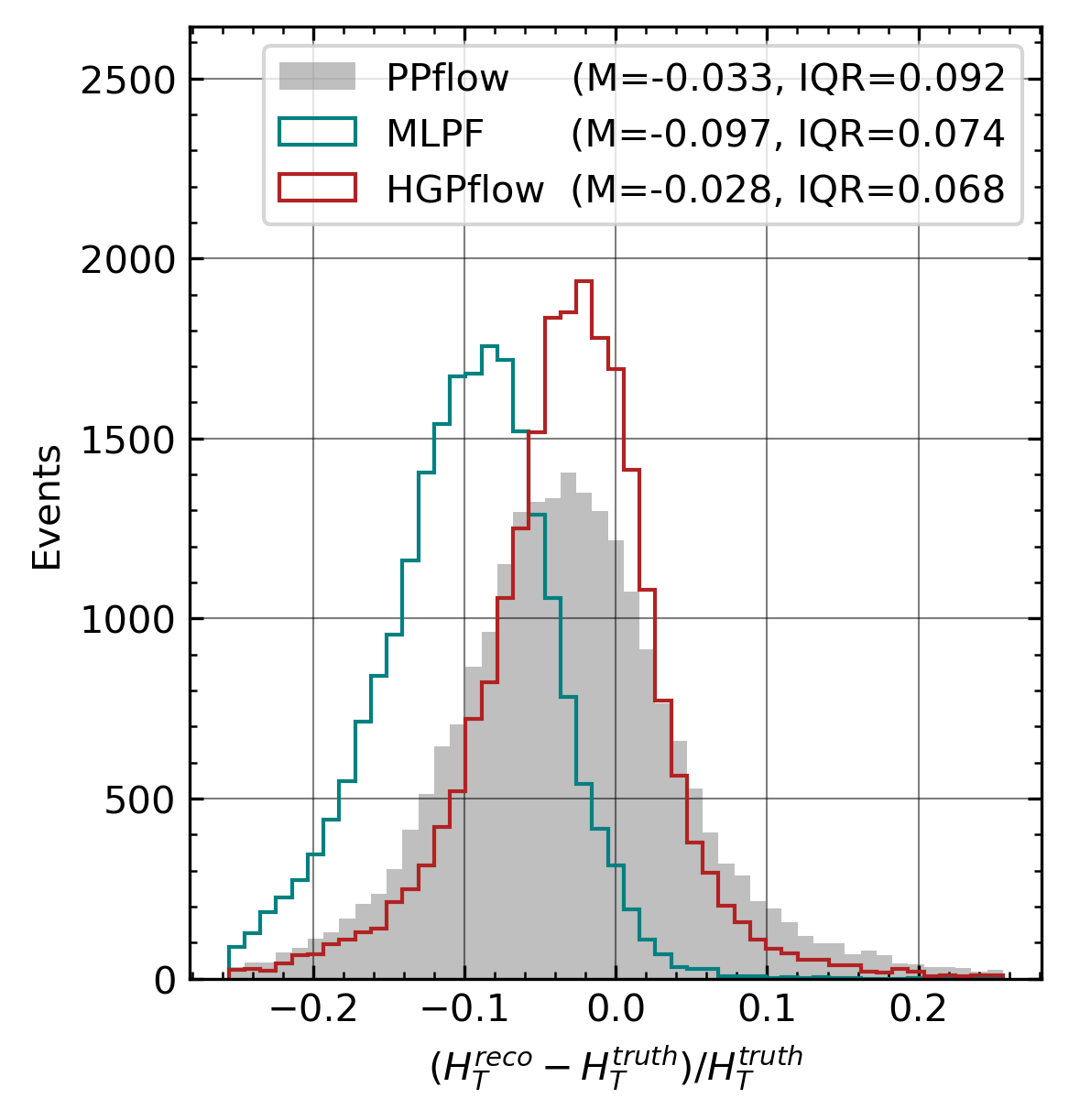}
        \caption{}
        \label{fig:ttbar_ht_residual}
    \end{subfigure}
    \caption{Residual distributions of (a) the vector \met and (b) scalar $H_\mathrm{T}$ sums of transverse momenta in full \ttbar events.}
    \label{fig:event_level_perf}
\end{figure*}

\subsection{Inclusive \ttbar sample (COCOA)}

Next, we evaluate performance for event-level observables and boosted topology. In Fig.~\ref{fig:event_level_perf} we check the performance of \hg for the \ttbar sample in terms of the vector and scalar sums of particle transverse momenta: $p_\mathrm{T}^\mathrm{miss} = |\sum_{i} \vec{p_\mathrm{T}}^i|$ and $H_\mathrm{T} = \sum_{i} |\vec{p_\mathrm{T}}^i|$. The superior residual distribution for \hg demonstrates not only that its partition-level training can extend to event-level inference, but that this also works for an event topology lying outside the distribution encountered during training. 

\begin{figure*}[!hb]
    \centering
    \begin{subfigure}[c]{0.32\textwidth}
        \includegraphics[width=\textwidth]{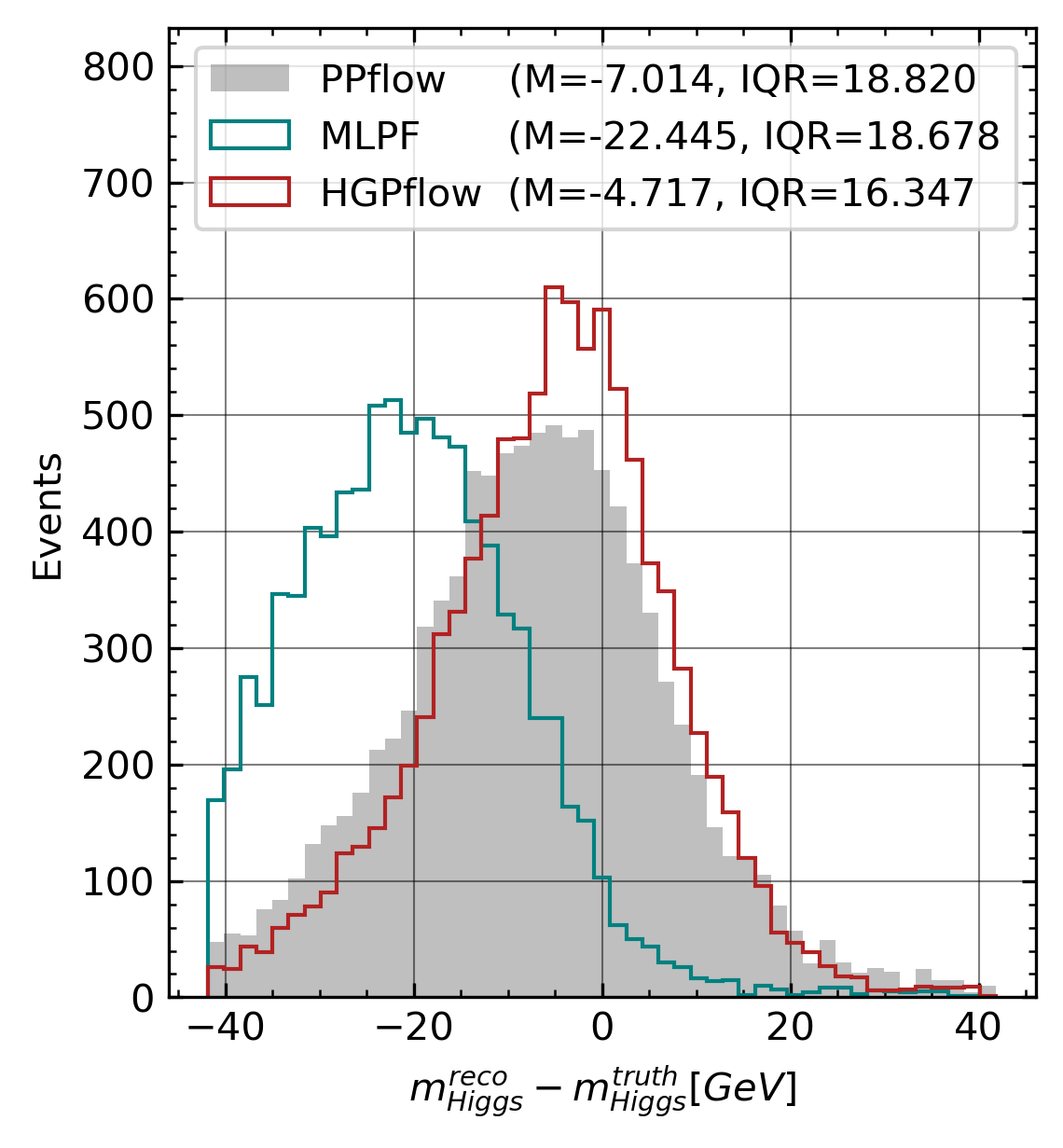}
        \caption{}
        \label{fig:zhbb_mass_residual}
    \end{subfigure}
    \begin{subfigure}[c]{0.335\textwidth}
        \includegraphics[width=\textwidth]{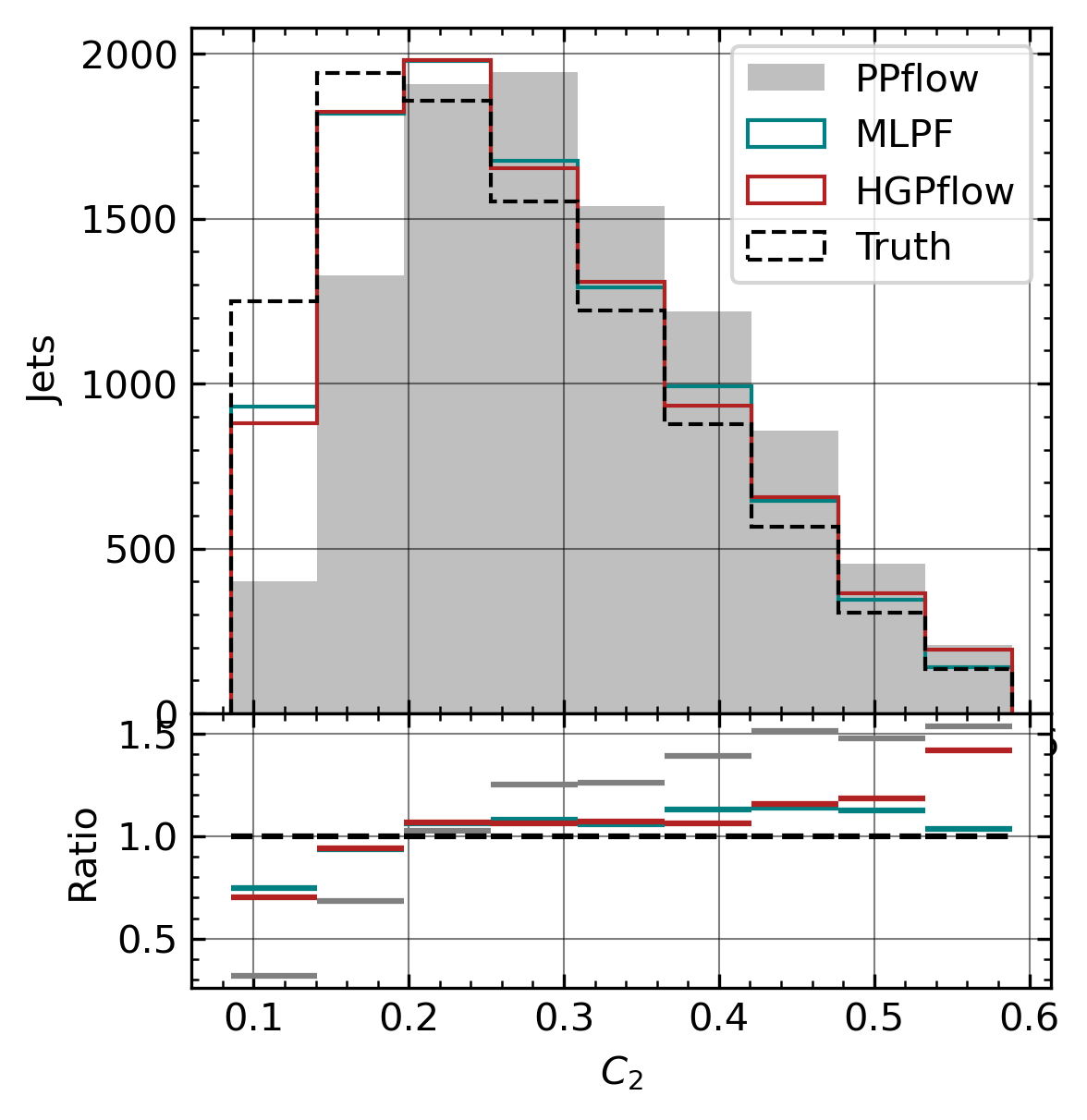}
        \caption{}
        \label{fig:zhbb_c2}
    \end{subfigure}
    \begin{subfigure}[c]{0.335\textwidth}
        \includegraphics[width=\textwidth]{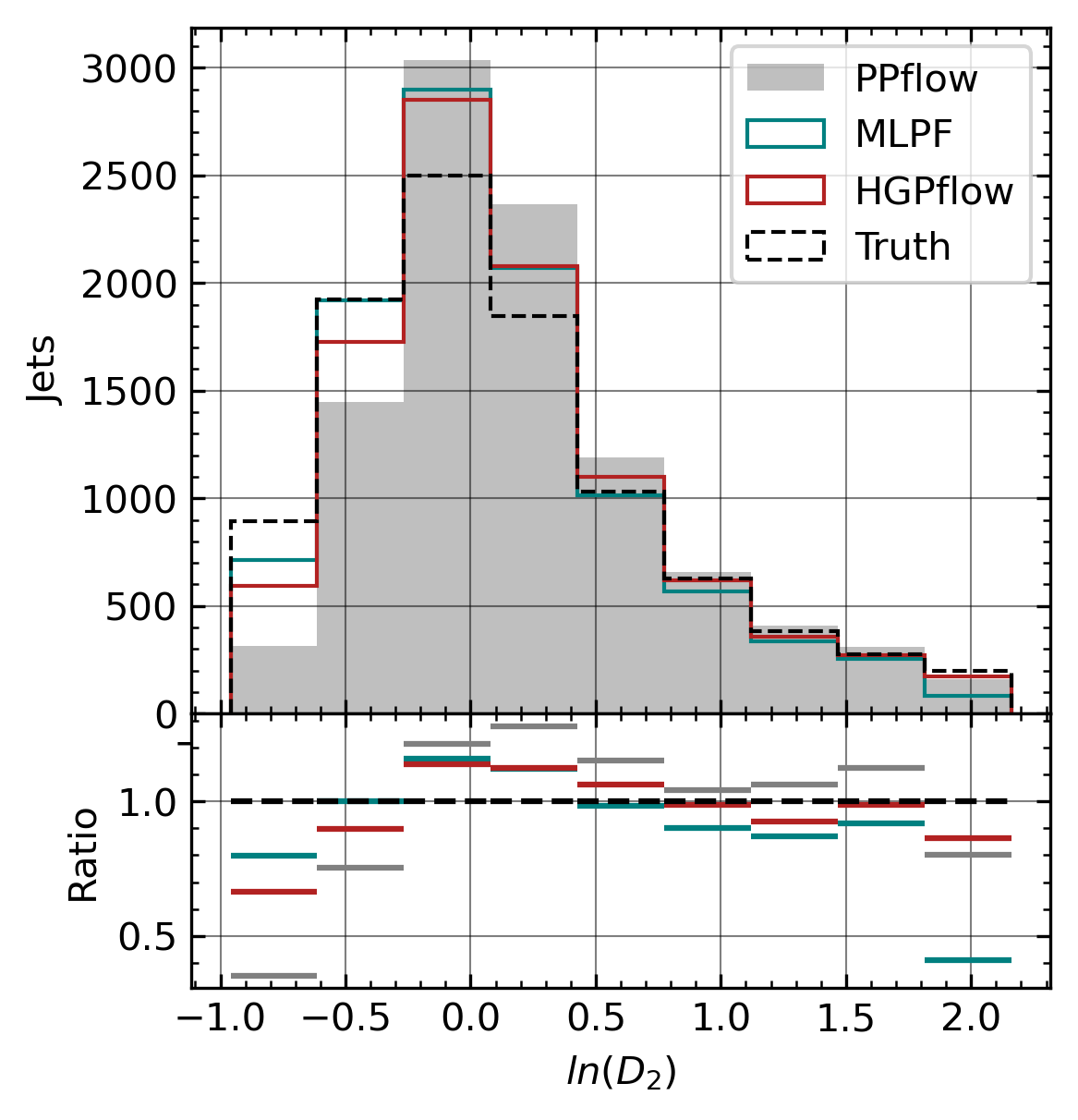}
        \caption{}
        \label{fig:zhbb_lnd2}
    \end{subfigure}
    \caption{(a) Residual distribution between the reconstructed Higgs mass, as the mass of the leading large-radius jet, and jet substructure variables (b) $C_2$ and (c) $ln(D_2)$ computed using truth particles and reconstructed particles in $Z(\nu \bar{\nu}) H (b \bar{b}))$ events.}
    \label{fig:higgs_substructure}
\end{figure*}

\subsection{Boosted $Z(\nu\nu)H(b\overline{b})$ sample (COCOA)}

We test how well the \hg training extends to out-of-distribution events containing boosted two-prong Higgs decays. After pooling the partitions for each event, anti-$k_\mathrm{T}$ jet clustering is run with $R=1.0$. Figure~\ref{fig:higgs_substructure}a shows the mass of the leading large-radius jet formed from \pf and \hg predictions relative to the truth jet in each event. As before, the dijet-trained \hg outperforms the benchmark \pf, demonstrating its ability to generalize to this boosted topology and reconstruct the Higgs mass. To further probe the ability of \hg to capture the two-prong nature of the system, we compute the substructure variables $C_2$~\cite{Larkoski:2013eya} and $D_2$~\cite{Larkoski:2014gra} using the constituents of the leading large-$R$ jet. The results are shown in Fig.~\ref{fig:higgs_substructure}b and \ref{fig:higgs_substructure}c, where it can be seen that the substructure distributions of \hg and MLPF resemble those of truth jets more closely than \pf, without either having a clear lead. 

\subsection{Dijet sample (CLIC) -- physics performance}\label{sec:dijet_clic}

While the various COCOA samples above demonstrate the potential of HGPflow to scenarios resembling that of the LHC (without pileup), here we explore the potential for application to $e^+ e^-$ collider experiments, represented by the CLIC detector simulation. Jets are created and matched in the same way as described in Sec.~\ref{sec:dijet_cocoa}, but for CLIC we use the generalized $k_\mathrm{T}$ algorithm with a radius parameter of 0.7. 

\begin{figure*}[!ht]
    \centering
    \begin{subfigure}[c]{0.33\textwidth}
        \includegraphics[width=\textwidth]{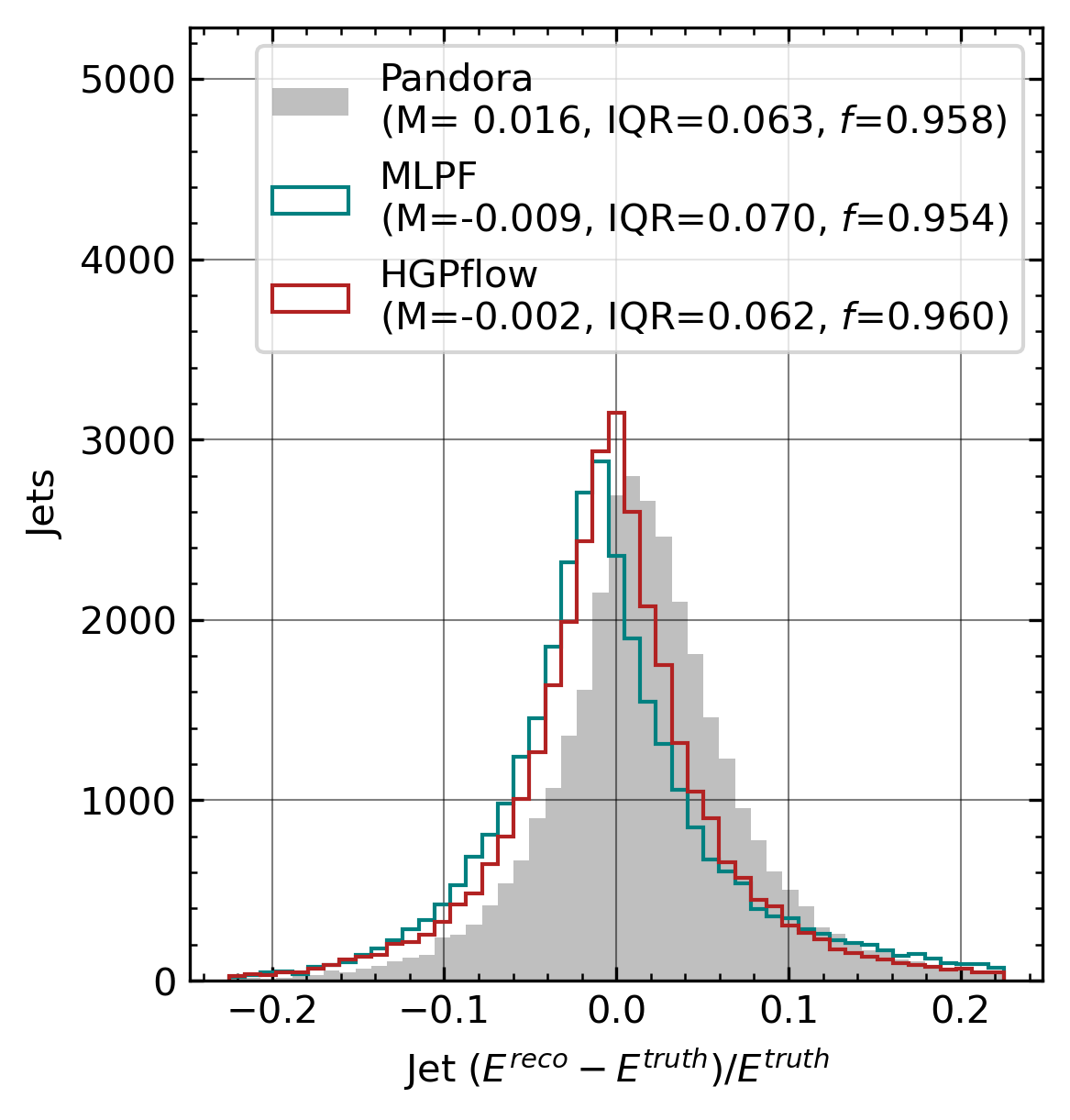}
        \caption{}
        \label{fig:clic_jet_res_e}
    \end{subfigure}
    \begin{subfigure}[c]{0.33\textwidth}
        \includegraphics[width=\textwidth]{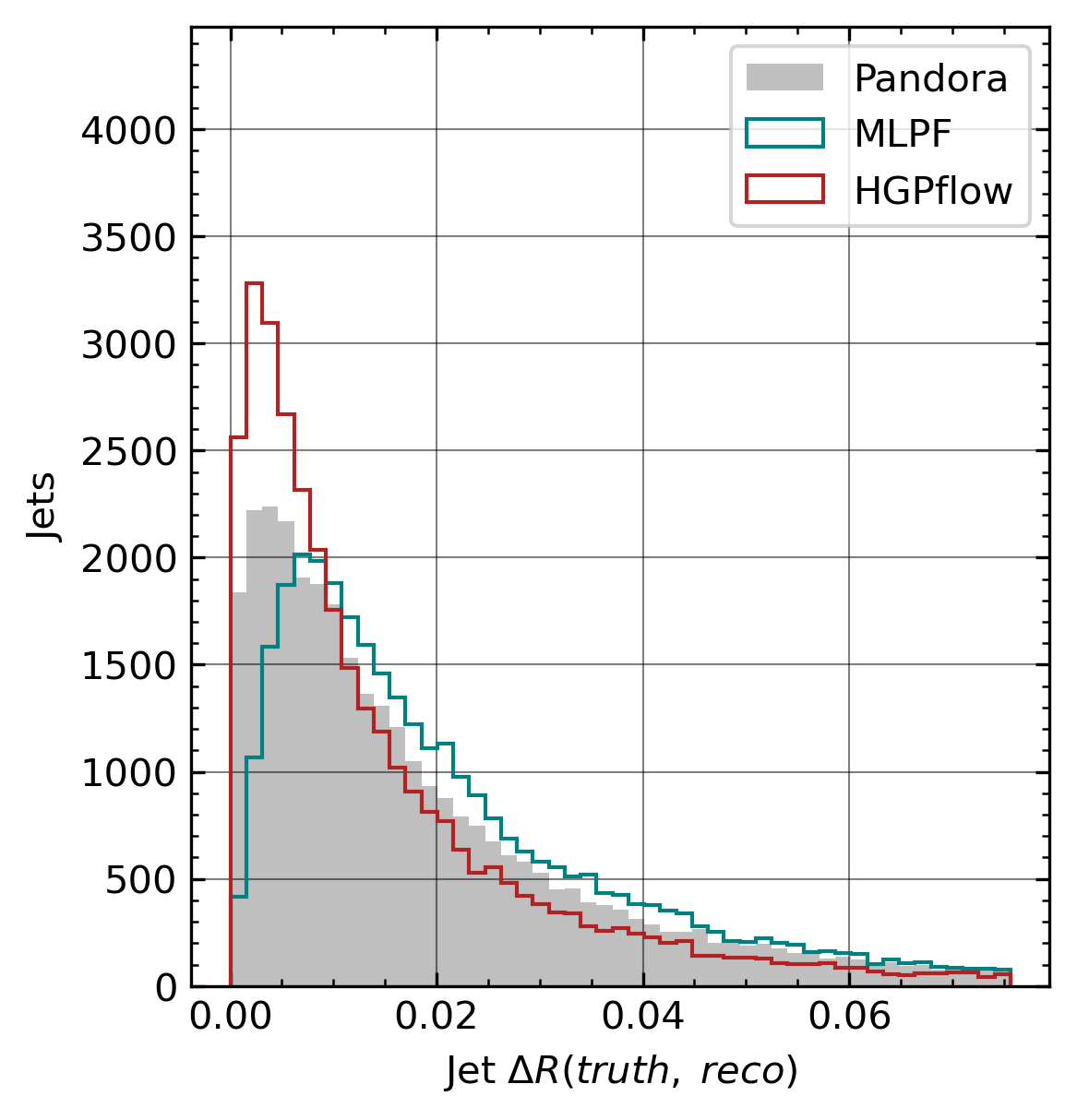}
        \caption{}
        \label{fig:clic_jet_res_dr}
    \end{subfigure}
    \begin{subfigure}[c]{0.33\textwidth}
        \includegraphics[width=\textwidth]{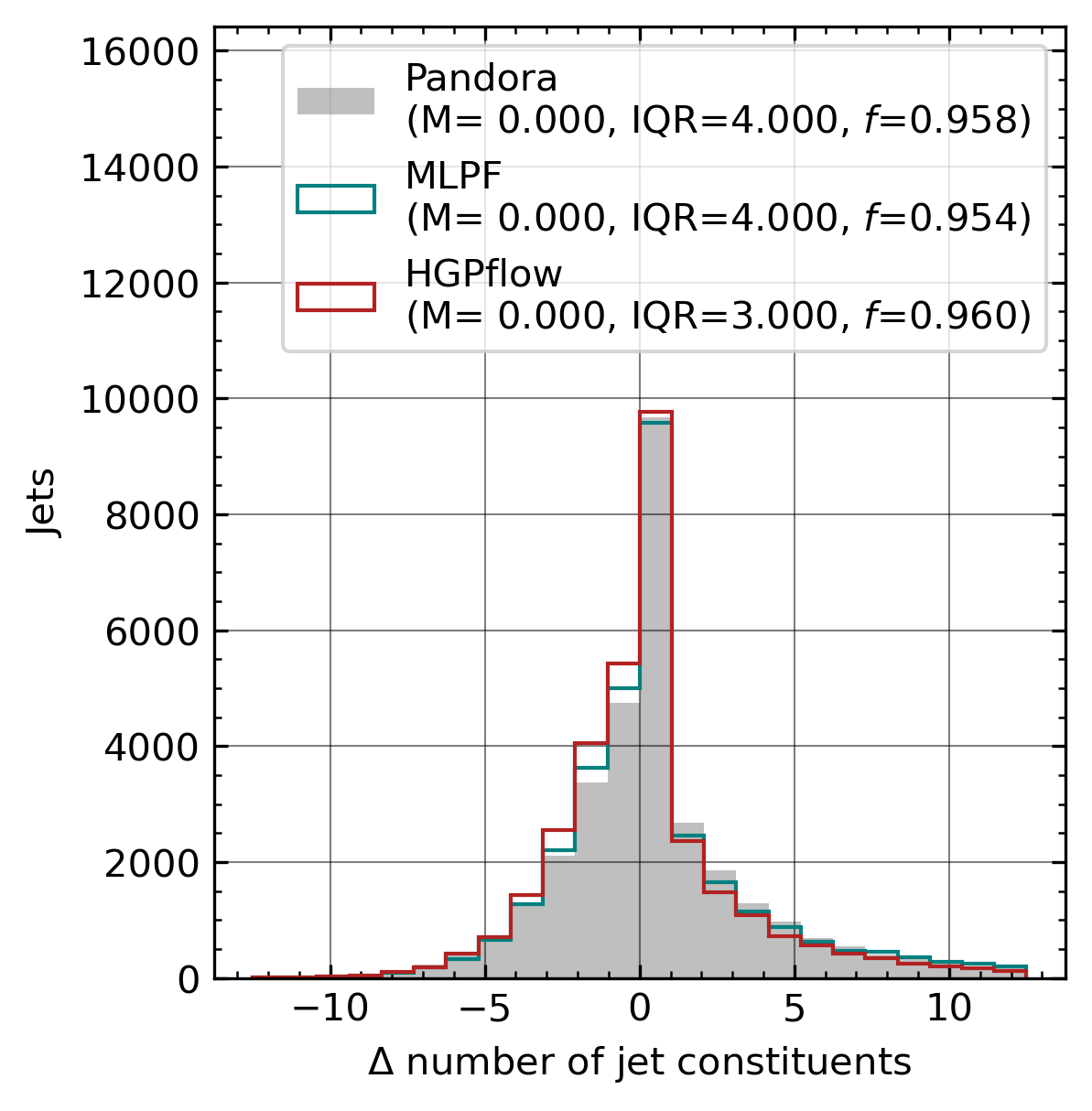}
        \caption{}
        \label{fig:clic_jet_res_nconst}
    \end{subfigure}
    \caption{Matching characteristics between truth and reconstructed jets in terms of (a) jet energy, (b) angular separation $\Delta R$, and (c) number of constituents $N_\mathrm{const.}$ for dijet events simulated in the CLIC detector. Up to two leading truth jets with $p_\mathrm{T} > 10$~GeV are taken from each event and matched to reconstructed jets based on $\Delta R$ up to a maximum of 0.1. The metrics M and IQR refer to the median and interquartile range, respectively, while $f$ is the efficiency of matching truth jets.}
    \label{fig:clic_leading_jet_residuals}
\end{figure*}

\begin{figure*}[!hb]
    \centering 
    \begin{subfigure}[c]{0.48\textwidth}
        \includegraphics[width=\textwidth]{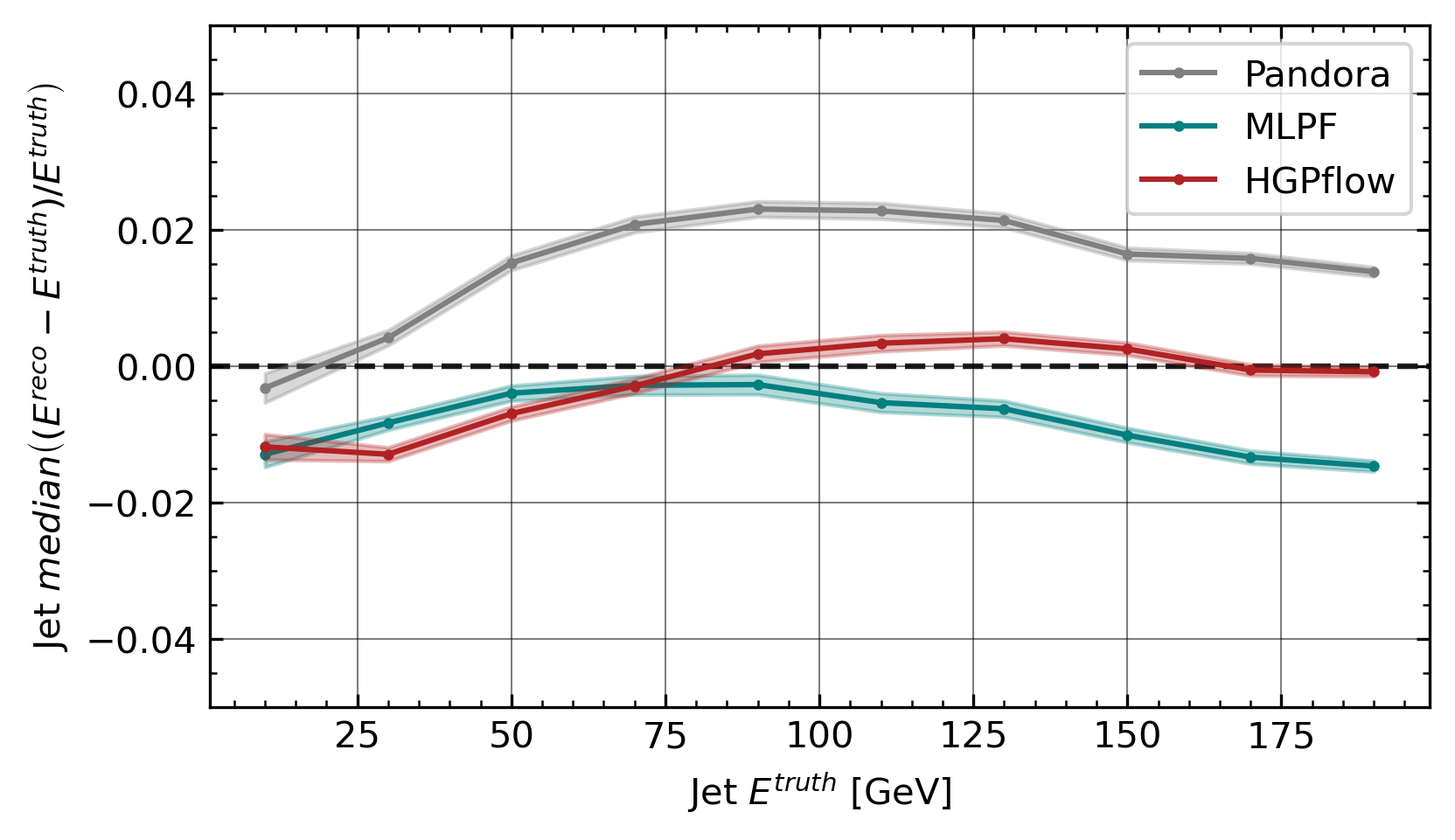}
        \caption{}
        \label{fig:clic_jet_res_median}
    \end{subfigure}
    \quad
    \begin{subfigure}[c]{0.48\textwidth}
        \includegraphics[width=\textwidth]{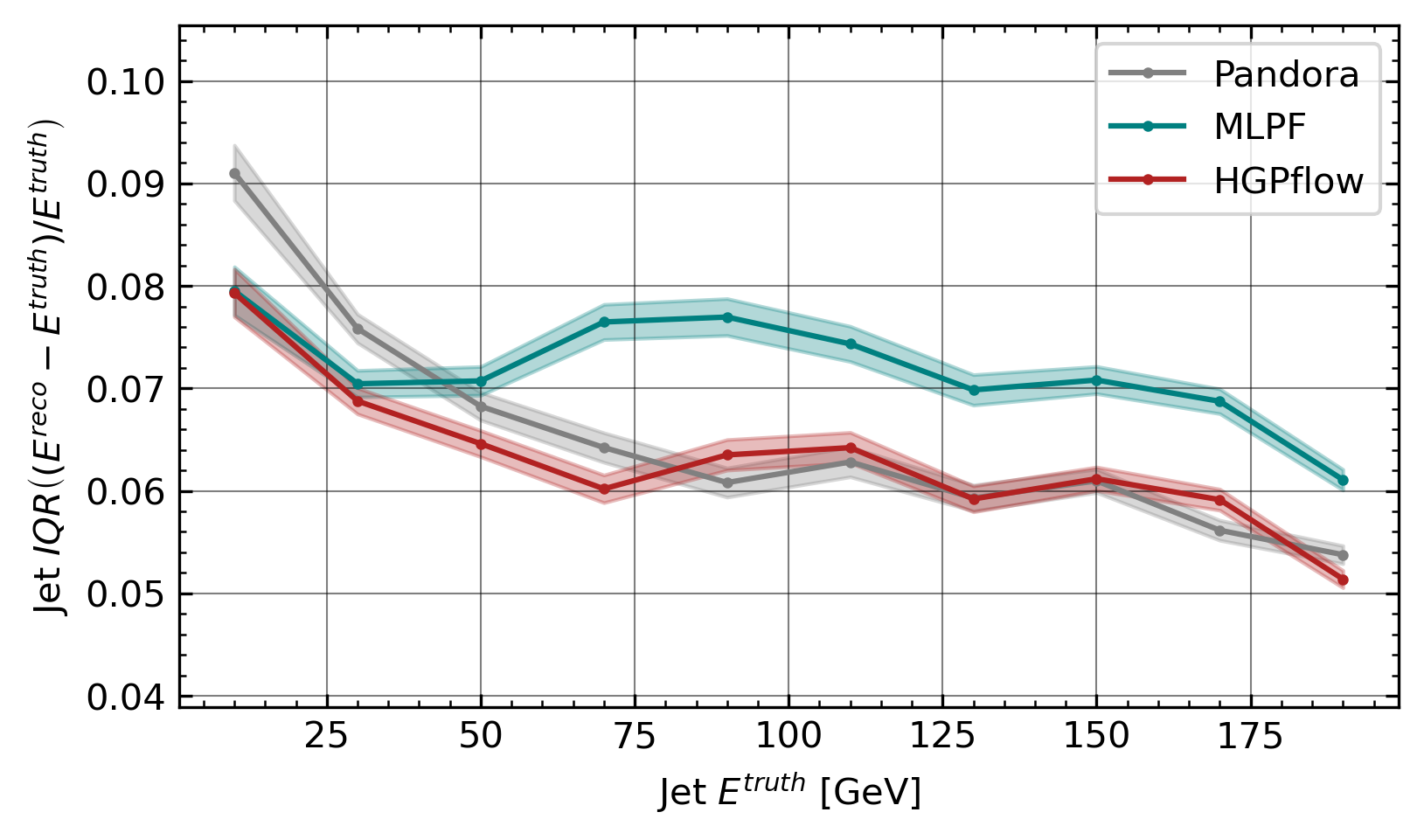}
        \caption{}
        \label{fig:clic_jet_res_iqr}
    \end{subfigure}
    \caption{(a) Median and (b) interquartile range of the jet energy relative residual distribution in bins of truth jet energy. Up to two leading truth jets with $p_\mathrm{T} > 10$~GeV are taken from each event and matched to reconstructed jets based on $\Delta R$ up to a maximum of 0.1.}
    \label{fig:clic_jet_resolution}
\end{figure*}

Figure~\ref{fig:clic_leading_jet_residuals} shows the performance of HGPflow on an independent set of dijet events in CLIC, with \pd and MLPF also shown for comparison. In terms of inclusive jet energy resolution, HGPflow performs slightly better than \pd, whereas the MLPF residual distribution has a larger scale and width. HGPflow outperforms both alternatives in its predictions of jet angular coordinates and the number of jet constituents.


In Fig.~\ref{fig:clic_jet_resolution}, we show how the median and interquartile range (IQR) of the jet energy relative residual distribution vary with truth jet energy. All three algorithms produce distributions centered within roughly $\pm 2\%$, with Pandora tending to overestimate and HGPflow being the most centered overall. For IQR, HGPflow outperforms Pandora by 0.5\%-1.0\% (absolute) for truth jets below 80 GeV, with the two performing similarly within error\footnote{The $\pm 1 \sigma$ error bands are computed under the assumption of normal distributions using $\sigma_\mathrm{M} = 0.93\cdot \mathrm{IQR} / \sqrt{N}$ and $\sigma_\mathrm{IQR} = 1.16\cdot \mathrm{IQR} / \sqrt{N}$.} above that. The IQR values of MLPF are similar to HGPflow until 40 GeV but worsen by 0.5\%-1.5\% (absolute) afterward.

\subsection{Dijet sample (CLIC) -- locality study}\label{sec:locality}

To investigate the locality of learned features in both models, we perturb one half of the dijet system and check the impact on the reconstruction of the other half. We start by selecting events from the test dataset with two leading central jets formed from calorimeter clusters using the generalized $k_\mathrm{T}$ algorithm with a radius parameter of 0.7. The $\Delta R$ between the jets is peaked at $\pi$ and is required to be above 1.5. In each event, the energies of clusters associated with the subleading jet are reduced by 80\%. This perturbation is clearly unphysical, serving merely as a stress test for potential long-range correlations in the model predictions. Finally, \hg and MLPF are used to reconstruct both the perturbed events and their original counterparts and compare the differences\footnote{Note that the random seeds are equal in the two evaluations to rule out differences from stochastic elements.}. 

\begin{figure}[h]
    \centering
    \includegraphics[width=0.33\textwidth]{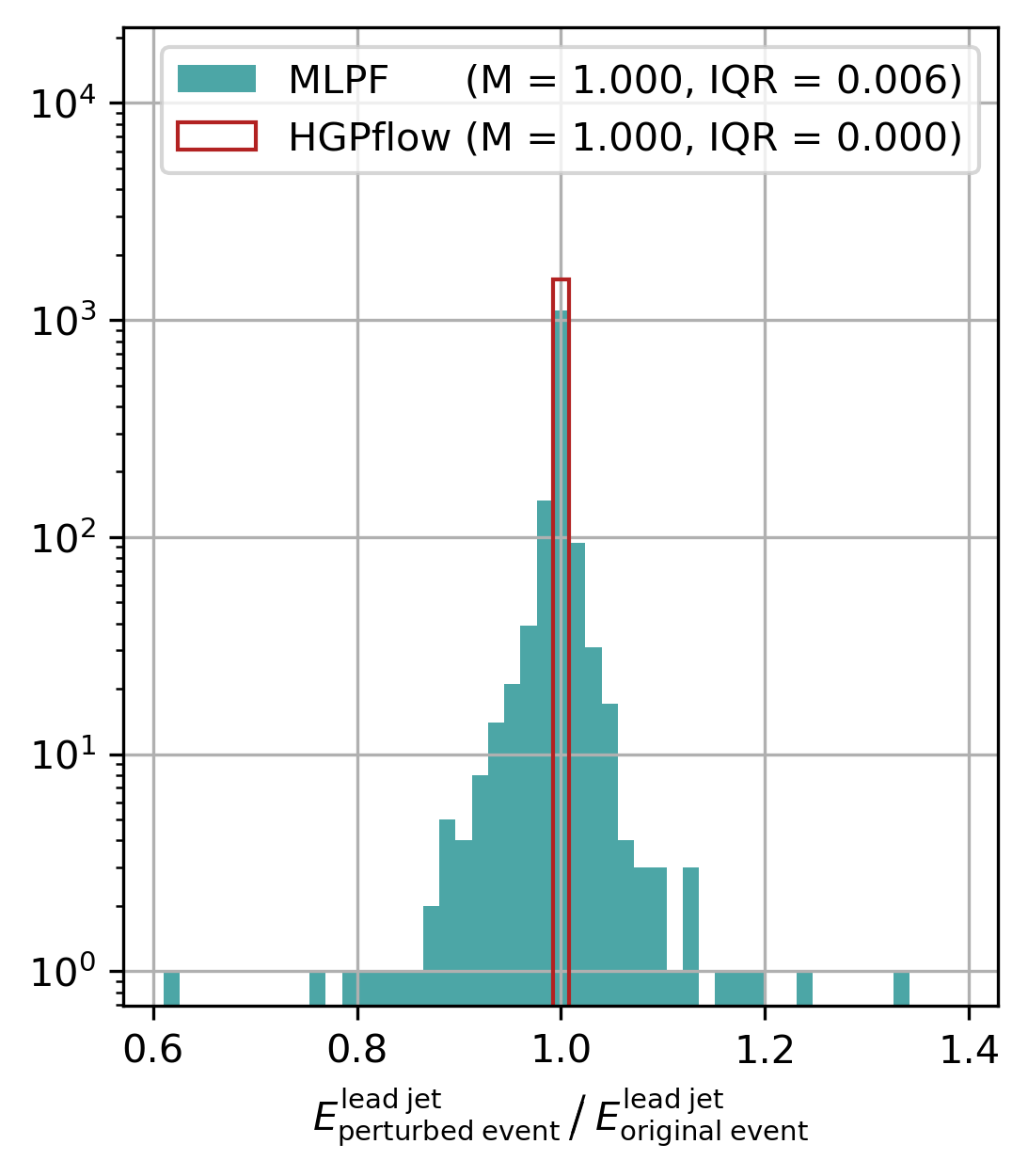}
    \caption{Test of the locality of the model predictions in dijet events simulated in CLIC. The ratio quantifies the dependence of leading jet energy to subleading jet energy when a perturbation of $-80\%$ is applied to the latter.}
    \label{fig:locality}
\end{figure}

Figure~\ref{fig:locality} shows the ratio of leading jet energy in the perturbed event relative to the original one. The leading jet should not be affected by the perturbation applied to the subleading jet due to their separation in $\Delta R$. This is realized for \hg, where event partitioning has broken this long-range correlation. In comparison, the test reveals a small but definite correlation between the jets reconstructed by MLPF. The limited impact of this correlation can potentially be credited to the approximate locality introduced by LSH.

\section{Discussion}\label{sec:discussion}
\begin{table*}[!htpb]
\centering
\begin{tabular}{ccccccccccccc}

\toprule

& \multirow{2}{*}{Model} & \# & Hyperparam. & \multicolumn{3}{c}{Jet $p_\mathrm{T} (E)$ residual (dijet)} & \multicolumn{2}{c}{$p_T^{miss}$ residual ($t\bar{t}$)} & \multicolumn{2}{c}{$H_T$ residual ($t\bar{t}$)} & \multicolumn{2}{c}{$m_{Higgs}$ residual}\\
& & param. & optimized & M & IQR & $f$ & M & IQR & M & IQR & M & IQR \\

\midrule

\multirow{3}{*}[-0.5em]{\rotatebox[origin=c]{90}{COCOA}} 
& PPFlow & -- & -- & -0.071 & 0.183 & 96.4\% & 0.047 & 0.658 & -0.033 & 0.092 & -7.014 & 18.820 \\ [0.5em]
& MLPF & 6.0M & no & -0.0079 & 0.149 & 96.9\% & -0.045 & 0.480 & -0.097 & 0.074 & -22.445 & 18.678 \\ [0.5em]
& HGPflow & \textbf{2.0M} & no & \textbf{-0.025} & \textbf{0.129} & \textbf{97.0\%} & \textbf{0.031} &\textbf{0.476} & \textbf{-0.028} & \textbf{0.068} & \textbf{-4.717} & \textbf{16.347} \\ [0.3em]
\midrule

\multirow{3}{*}[-0.5em]{\rotatebox[origin=c]{90}{CLIC}} 
& Pandora & -- & -- & (0.016) & (0.063) & 95.8\% & -- & -- & -- & -- & -- & -- \\ [0.5em]
& MLPF & 6.0M & yes* & (-0.009) & (0.070) & 95.4\% & -- & -- & -- & -- & -- & -- \\ [0.5em]
& HGPflow & \textbf{5.2M} & no & (\textbf{-0.002}) & (\textbf{0.062}) & \textbf{96.0\%} & -- & -- & -- & -- & -- & -- \\ [0.3em]

\bottomrule

\multicolumn{7}{l}{\raggedleft
\textit{*performed using the ground truth definition of \cite{Pata:2023rhh}.}
}
\end{tabular}
\caption{Performance summary of all the algorithms on both the COCOA and CLIC datasets in terms of median (M) and interquartile range (IQR) of the residual distributions and jet matching efficiency~$f$.}
\label{tab:perf_summ}
\end{table*}

Our findings indicate that \hg coupled with event partitioning is a promising solution for particle reconstruction at existing LHC detectors and potential future detectors, represented by CLIC. Despite having access to only one partition at a time, \hg scales well to full events for different physics processes, as expected of a local reconstruction model. In the following discussion, we summarize the performance comparison with MLPF and note important ways in which our work in this paper can be extended in the future.

The performance of \hg and MLPF on the various samples considered is summarized in Tab.~\ref{tab:perf_summ}. Unlike MLPF, HGPflow is trained to disentangle cases where a single calorimeter cluster contains energy from more than one particle. This edge appears to be most relevant for lower granularity calorimeters, as suggested by comparing the relative performance of HGPflow and MLPF in modeling the number of jet constituents in COCOA (Fig.~\ref{fig:cocoa_jet_res_nconst}) versus CLIC (Fig.~\ref{fig:clic_jet_res_nconst}).

We have focused on a relatively simple definition of the region on which particle reconstruction is considered to be ``local''. More detailed investigation is needed to better quantify this definition and how it impacts a model's ability to function in a generic way. The potential for bias from other sources should also be considered. For example, the effect of fragmentation on jet energy distribution likely influences our current training by introducing sensitivity to a specific physics model, in this case, that of \textsc{Pythia}. This challenge is not unique to particle reconstruction, and may also benefit from domain adaptation techniques \cite{CMS-DP-2024-063,Sheldon:2024sbe,Baalouch:2019fhm}. 

Another strategy to suppress dependence on specific physics information comes through the design of the training sample. Training on a dataset that contains diverse physics processes can discourage the network from over-optimizing on any single process, as was done in \cite{Pata:2023rhh}, for example. A potential shortcoming of this approach is that it is not obvious how to optimize the composition to minimize sample dependence, particularly when accounting for physics processes beyond the Standard Model.

Perhaps the most robust approach would be to train on artificial samples created by overlaying multiple single particles. In this case, the type, kinematics, and isolation of particles could be fully configurable while having no dependence on generator modeling or physics process. It would need to be shown, however, that a model trained in this way could generalize well to samples of real physics processes.

At the very least, we envision that a more optimized training dataset will be included in future work, covering a larger kinematic range and augmented with single-particle samples. This would enable the exploration of \hg performance for isolated leptons and photons, which have so far been omitted due to a lack of training examples. It would also present an opportunity to develop a fiducial particle definition for electrons and photons that accounts for bremsstrahlung and photon conversions (see Sec.~\ref{sec:fiducial}). A more complete definition might also feature separate classes for nontrivial decay signatures such as neutral pions and tau leptons (see \cite{Tani:2024qzm}).

The impact of pileup is similarly left to future work. This question is very important for gauging how well \hg could perform in realistic collision environments at future runs at the LHC. If the model trained without pileup struggles to generalize well, a dedicated approach to pileup suppression might be incorporated as a classification problem, or at least as a global conditioning variable.

The overall excellent performance of \hg suggests that edge effects from event partitioning are manageable. It was noted in Sec.~\ref{sec:MS} that less than 10\% of the total energy in dijet events arise from hadronic showers that get split between two MS clusters, i.e., residual particles. However, the relatively large rate of unmatched neutral hadrons observed in Fig.~\ref{fig:eff_fr_purity} might be driven partly by residual particles.
Future work could thus further quantify and reduce this impact by optimizing the bandwidth of MS clustering, refining the distance metric, and testing alternative algorithms. Provided a suitable clustering objective (e.g. minimizing cluster size and edge effects simultaneously), a partitioning approach based on DL may prove to be a versatile solution. Additionally, making the partitioning step GPU compatible could lead to significant speed-up and enable running it ``on the fly'' during model inference.

\section{Conclusions}
In this paper, we extended the paradigm of particle reconstruction with hypergraph learning (\hg) to full collision events. Instead of operating directly on the full event, we used a clustering algorithm to define smaller partitions on which to train and evaluate \hg. In addition to reducing memory usage, this approach imposes locality on the features extracted by the model, avoiding potential bias from long-range correlations. We demonstrate that \hg provides excellent modeling of particles, jets, and global observables that meets and, in most cases, exceeds that of its parametric and deep learning counterparts. This progress underscores the potential of deep learning to enhance particle reconstruction at current and future colliders while identifying key areas for further exploration in this domain.

\section{Code and data availability}

The code for event partitioning, the HGPflow model, and performance evaluation are available at \\
\url{https://github.com/nilotpal09/HGPflow} \\
along with the trained model parameters. The code for training MLPF on CLIC data with our modified target and for training on COCOA events is available at \\
\url{https://github.com/annaivina/particleflow-fork} \\
along with the trained model parameters in each case. The COCOA samples are available at \cite{data_zenodo}, and the CLIC samples were published previously in \cite{pata_zenodo}.

\section{Acknowledgements}
NK, ED, AI, and EG are supported by the BSF-NSF Grant No. 2020780 and the Weizmann Institute and MBZUAI Collaboration Grant. LH is supported by the Excellence Cluster ORIGINS, which is funded by the Deutsche Forschungsgemeinschaft (DFG, German Research Foundation) under Germany’s Excellence Strategy - EXC-2094-390783311.

\bibliography{hgpflow}
\bibliographystyle{unsrt}

\clearpage
\end{document}